\documentclass[letterpaper,final,conference,twocolumn,10pt,table]{IEEEtran}

\IEEEoverridecommandlockouts

\usepackage{graphicx}
\usepackage{balance} %
\usepackage{paralist}
\usepackage[inline]{enumitem}%
\usepackage{pifont}%
\usepackage{diagbox}
\usepackage{layout}
\usepackage[english]{babel}
\usepackage{graphicx}
\usepackage[rightcaption]{sidecap}
\usepackage{cite}
\usepackage{amsmath,amssymb,amsfonts}
\usepackage{algorithmic}

\usepackage[table]{xcolor}

\usepackage{multirow}
\usepackage{subcaption}
\usepackage{tabularx}
\usepackage{sidecap}
\usepackage{array}
\usepackage{textcomp}
\usepackage{wrapfig}
\usepackage[normalem]{ulem}
\usepackage{caption}

\usepackage{diagbox}

\usepackage{mathtools}
\usepackage{url}                %

\usepackage{rotating}
\usepackage{lscape} 
\usepackage{afterpage}
\usepackage[htt]{hyphenat}

\usepackage[]{hyperref}         %
\hypersetup{                    %
  colorlinks=true,
  linkcolor={green!80!black},
  citecolor={red!70!black},
  urlcolor={blue!70!black}
}

\def\displaytoconsider{0}
\ifnum\displaytoconsider=1
\newcommand{\toconsider}[1]{{\color{red} #1}}
\else 
\newcommand{\toconsider}[1]{\ignorespaces}
\fi

 \def\showcomments{1}
\ifnum\showcomments=1
    \usepackage[textsize=small,textwidth=0.6in]{todonotes}

    \newcommand{\David}[2][]{\todo[color=blue!30,#1]{David: #2}}
    \newcommand{\Shawn}[2][]{\todo[color=orange!30,#1]{#2}}
    \newcommand{\Viau}[2][]{\todo[color=green!30,#1]{Viau: #2}}

\else
    \newcommand{\David}[2][]{\ignorespaces}
    \newcommand{\Shawn}[2][]{\ignorespaces}
    \newcommand{\Viau}[2][]{\ignorespaces}

    \newcommand{\todo}[1]{\ignorespaces}
\fi

\def\BibTeX{{\rm B\kern-.05em{\sc i\kern-.025em b}\kern-.08em
    T\kern-.1667em\lower.7ex\hbox{E}\kern-.125emX}}

\newcolumntype{P}[1]{>{\centering\arraybackslash}p{#1}}

\newcommand{\subhead}[1]{\vspace{0.5mm} \noindent{\textbf{#1.}}}

\newcounter{rowcount}

\usepackage{graphicx} %

\title{\cz{}: Certifying Web Page Interactions with Computer Vision}

\author{\IEEEauthorblockN{He Shuang}

\IEEEauthorblockA{
\textit{University of Toronto}\\
he.shuang@mail.utoronto.ca
}
\and
\IEEEauthorblockN{Lianying Zhao}
\IEEEauthorblockA{
\textit{Carleton University} \\
lianying.zhao@carleton.ca
}
\and
\IEEEauthorblockN{David Lie}
\IEEEauthorblockA{
\textit{University of Toronto}\\
david.lie@utoronto.ca
}
}

\def\cz{vWitness}
\def\Vspecs{{\bf \rm V\kern-.05em{\sc S\kern-.08em p\kern-.08em e\kern-.08em c}s}}
\newcommand{\Vspec}{{\rm V\kern-.05em{\sc S\kern-.08em p\kern-.08em e\kern-.08em c}}}

\begin{document}
\pagestyle{plain}

\maketitle

\begin{abstract}
Web servers service client requests, some of which might cause the web server to perform security-sensitive operations (e.g. money transfer, voting). An attacker may thus forge or maliciously manipulate such requests by compromising a web client.  Unfortunately, a web server has no way of knowing whether the client from which it receives a request has been compromised or not---current ``best practice'' defenses such as user authentication or network encryption cannot aid a server as they all assume web client integrity.

To address this shortcoming, we propose \cz{}, which ``witnesses'' the interactions of a user with a web page and certifies whether they match a specification provided by the web server, enabling the web server to know that the web request is user-intended.  The main challenge that \cz{} overcomes is that even benign clients introduce unpredictable variations in the way they render web pages.  \cz{} differentiates between these benign variations and malicious manipulation using computer vision, allowing it to certify to the web server that 1) the web page user interface is properly displayed 2) observed user interactions are used to construct the web request. Our \cz{} prototype achieves compatibility with modern web pages, is resilient to adversarial example attacks and is accurate and performant---\cz{} achieves 99.97\% accuracy and adds 197ms of overhead to the entire interaction session in the average case.

\end{abstract}

\section{Introduction}
\label{sec:intro}
Web applications are increasingly used to implement security-sensitive services such as financial transactions, access to social services and e-health. These services are sensitive because they allow a user to make important and in some cases, irrevocable actions that may affect the user's financial well-being, freedom and safety.  The security of these services rests on the assumption that requests leaving user clients are intended by the human user. This assumption depends on two factors: first, that we can correctly authenticate the principal making the request and second, that the integrity of the user-page interaction is maintained. 

While web authentication has been well-studied (e.g. passwords and other methods~\cite{dhamija2000deja,9152694,ross2005stronger,fingerprinting_authentication}), securing user-page interaction is an umbrella definition that covers three goals~\cite{VButton, protection}: the proper page rendering, the proper processing of user inputs and the proper construction of outgoing requests. \textit{Violating any one of the goals allows an attacker to send out user-unintended requests. }
The following are examples of user-page interaction attacks: clickjacking~\cite{clickhijacking} shows a deceptive user interface (UI) to trick a victim user into performing unintended actions, click fraud forges user inputs on web elements (e.g. clicks)  for monetary benefits and cross-site request forgery~\cite{khodayari2021jaw} crafts user-unintended requests to remote servers. 

Existing defenses for user-page interaction assume an unprivileged/remote attacker and rely on client-side enforcement such as browser-based flags (e.g. x frame options~\cite{XFrameOp32:online} for page overlaying) and policies (e.g. cross-origin resource sharing~\cite{CrossOri34:online}, content security policy~\cite{ContentS85:online}, sub-resource integrity~\cite{Subresou12:online} for unauthorized code execution). 
These defenses can be bypassed by client-side malware~\cite{agarwal2021first,BREXKILI31:online}.  For example, Scranos~\cite{scranos}, a web-targeting client-side rootkit, attacks remote servers by forging user interactions, chaining into sophisticated attacks (e.g. installing malicious browser plugins, stealing credentials and cookies, and spamming messages). Further, leveraging its privilege, it can disable in-browser security mechanisms and install additional payloads to enrich its capabilities. 
What is special about Scranos is that 1) its forged requests are indistinguishable from user-intended ones as observed by the server and 2) it is privileged and can bypass unprivileged defenses (e.g. browser-based). 

Before Scranos, the research community attempted to deal with a hypothetical \textit{privileged} adversary. 
Solutions such as NAB~\cite{Not-a-bot}, can ensure that all requests are generated by I/O from a human user, but they cannot semantically distinguish whether the human-generated I/O causally resulted in the request.  Alternatively, other works implement high-assurance, small trusted computing base (TCB) clients that are resistant to subversion, but suffer from limited functionality: 
Vbutton~\cite{VButton}, Fidelius~\cite{Fidelius} and ProtectiON~\cite{protection} pack web page rendering (i.e. renderer), input processing (i.e. device drivers) and request construction (i.e. JavaScript engine) into trusted execution environments (TEE, e.g. ARM TrustZone, Intel SGX). But to keep the TCB small, the functionality support is extremely limited. For instance, Fidelius only supports rendering textboxes and handling keyboard inputs (i.e. no mouse support). These limitations can be exploited by UI attacks~\cite{context_hiding_attack} (due to the coexistence of trusted and untrusted UI elements) and limit the compatibility of such TEE-based defenses (i.e. almost no modern web page is textbox-only). 

In this paper, we propose an approach that works for fully-functional commodity web browsers by certifying to a remote server that the interactions between a user and a web page match some server-supplied specifications under an OS-level privilege malware. 
Instead of trying to secure the entire software stack that the user interacts with, our proposal, which we call \cz{} (short for ``virtual witness''. We take inspiration from how the signing of some legal documents must be ``witnessed'' by another party to be considered valid.) is a trusted component that passively observes the UI interactions the user has with the web page through \textit{screenshots}. The screenshots, in our design, are samples of the display frame buffer taken from a lower-level and thus, are invisible to the client software stack (e.g. OS and browser) making it immune to tampering.

With the screenshots, \cz{} leverages computer vision, to
\begin{enumerate*}
\item ensure the web interface was correctly displayed to the user,
\item observe user-page interactions and ensure the final outgoing request is properly constructed based on correct semantics.
\end{enumerate*}
To achieve both goals, \cz{} needs server-provided specifications (called \Vspec{}) as web page appearances and request construction are both server-specific. As part of this work, we provide automatic \Vspec{} construction scripts. 

Requests satisfying both goals above achieve \textit{interaction integrity} and are considered to be \textit{user-intended} (when paired with authentication, but not in the scope of this work). \cz{} uses a cryptographic signature to convey whether a request satisfies interaction integrity to remote servers. 

\cz{}'s novel use of computer vision is to differentiate between  \textit{benign rendering variations} and \textit{malicious display manipulations} on the screenshots.  
A key challenge is that it is impossible for a web server to know, a priori, at the pixel level, exactly how a web page will be rendered on a user's device even in the absence of an attacker.
This is due to the client-side variation in the rendering stack which consists of browsers, drivers, operating systems (OS, e.g. available fonts), and configuration settings (e.g. ClearType~\cite{cleartype}). Such variations cause enough differences even enabling fingerprinting~\cite{pixel_perfect}.
\cz{}'s vision-based validator can tolerate rendering variations while still being more accurate than previous works using pixel-by-pixel comparison ~\cite{VButton} and image hash~\cite{image_hash_tampering}.
Computer vision is vulnerable to adversarial machine learning attacks, for which we devise and evaluate a set of \cz{}-specific mitigation (as opposed to general defenses) that increases the robustness by 5.14$~\times$. 
While validation-based (i.e. signatures) and isolation-based (i.e. TEE) secure systems each have their pros and cons, \cz{}'s limitations are those imposed by the use of computer vision, which we must be able to see to interpret it (e.g. \cz{} can not verify file uploads) and can become computationally expensive to verify content-rich UI elements (e.g. complex animations or videos). While we leave these as future work, we believe they do not prevent \cz{} from working with most security-sensitive pages.

\cz{}'s approach has unique performance traits: 
\begin{enumerate*} 
    \item \cz{} runs in the background and is not on the critical path of the user-page interaction: it introduces minimal overhead at interaction time.  
    \item \cz{}'s validation is concurrent to the user interaction, and thus, depending on the length of the session and the speed of validation, a delay is added to the final request when user-page interaction has ended. 
    \item \cz{}'s validation can be incremental and benefits from caching and a GPU. 
\end{enumerate*}

\cz{} certifies interaction integrity and conveys the result to the server using a single bit of information (i.e. whether interaction integrity satisfies or not). It does not reveal unnecessary details to the server as all processing of screenshots remains local on the user client and screenshots are erased after a session finishes.

\noindent{\textbf{Contributions}}. 
\begin{itemize}[nolistsep]
    \item We propose the idea of achieving interaction integrity through passive observation. We trained computer vision-based validators that can distinguish benign rendering variation and malicious UI tampering. 
    We implemented a client-side prototype, called \cz{}, in Xen's dom0, which is able to certify requests' interaction integrity to remote servers. 
    We constructed server-side scripts to automate \Vspec{} construction. 
    \item We propose four \cz{}-specific defenses against adversarial attacks that increase the robustness of \cz{}'s text model by a factor of 5.14 over a reference on five attacks from CleverHans~\cite{papernot2018cleverhans} and AutoAttack~\cite{croce2020reliable}.
    \item We evaluate \cz{}'s compatibility, performance and accuracy on two datasets: Clickbench~\cite{clickshield} and Jotform real-life web pages. The results show that \cz{} is compatible with 10$\times$ more pages than previous proposals, has an accuracy of 99.97\% when validating pages and interactions, and provides performance (197/230ms delay for GPU/CPU setups) better than TEE-based works. 
\end{itemize}

\cz{} source code is available at \url{github.com/dlgroupuoft/vWitness-DSN23}.

\section{Security Model} \label{sec:2}
With current web technologies, remote servers cannot determine if a request is intended by human users or if the request has been tampered with or forged by malware on the user client. Accepting malware-constructed/tampered requests can harm the service as well as the user (e.g. unintended Youtube subscription~\cite{scranos} and cryptocurrency loss\cite{ViperSof64:online}).

\subhead{Threat Model} We assume the adversary controls malware with OS-level privileges on the user client. The malware can tamper with any guest OS components or user software such as the browser, but not a hypervisor that \cz{} relies on for its integrity. Thus, firmware- and hypervisor-level malware is considered out-of-scope. This threat model is in line with previous works~\cite{VButton, protection, Fidelius} as well as the assumption on hypervisor integrity~\cite{chen2008overshadow}.

\cz{} considers phishing attacks to be out of scope as our defense requires server integrity and cooperation. In addition, \cz{} does not address the authentication problem as it is orthogonal to interaction integrity, and \cz{} considers attacks on user confidentiality to be out of scope (e.g. stealing sensitive inputs). We do not consider availability or denial of (\cz{}) service attacks as blocking \cz{} does not prevent users from using the unprotected page, which will eventually lead to rejected requests due to missing \cz{} certification. Finally, we assume an honest user who enters the inputs they intend into the web page (i.e. they don't try to later equivocate about their earlier actions).

\begin{table*}[t]
\centering
\begin{tabular}{p{1.8cm}p{4.9cm}p{5cm}p{4.5cm}}
\rowcolor{gray!75} Attack Vectors    & Description and Example                                     & Example Attack                                                                                                                     & Other Defenses                             \\
\rowcolor{gray!25} Request forgery   & Malware forges requests without user input                     & Fake subscription, spamming, unaware payments                                                                                     & Captcha                                     \\
Request tampering & Malware intercepts and tampers with user requests & Cryptocurrency transfer redirection~\cite{ViperSof64:online} & Third party detection~\cite{kapravelos2014hulk}, trusted request construction~\cite{protection, Fidelius} \\
\rowcolor{gray!25} User interface (UI) tampering  & Malware manipulates UI to confuse or deceive user (i.e. clickjacking~\cite{clickhijacking})               & Facebook like hijacking~\cite{InfoSecH86:online}, Tweetbomb~\cite{Explaini11:online}, Scarno~\cite{scranos}                                                                                                               & Server-side confirmation~\cite{WhydoIge64:online, TradingA12:online}, anomaly detection~\cite{juba2015principled}
\end{tabular}
\caption{Attack vectors.}
\label{tab:attack_vectors}
\vspace{-5mm}
\end{table*}

\begin{figure}[t]
  \begin{center}
       \includegraphics[width=0.95\linewidth,trim=0mm 1mm 0mm 0mm,clip]{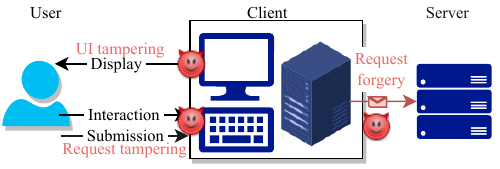}
  \end{center}
  \caption{Steps of user-page interaction and the attack vectors.}
  \label{fig:vectors}
  \vspace{-3mm}
\end{figure}

\begin{figure}[t]
    \includegraphics[width=0.98\linewidth]{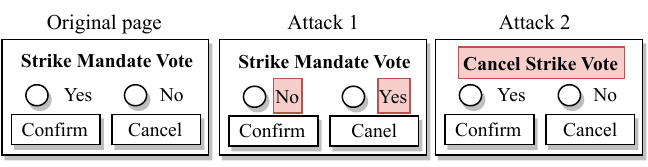}  
  \caption{UI tampering leads to unintended user actions and requests. In both attacks, only the displayed text are altered by malware. }
  \label{attack:context_hiding}
  \vspace{-5mm}
\end{figure}

\subhead{Attack Vectors} 
The attacker's goal is to submit a request to the web server and convince the web server that it came from a legitimate user. To achieve this, there are three attack vectors detailed in Table~\ref{tab:attack_vectors} summarizing what an attacker can do.  Some prior proposals~\cite{protection,Fidelius} use TEEs to protect the user client, thus relying on a small trusted computing base (TCB) to provide assurance. The small TCB only enables incomplete UI support making them vulnerable to tampering of the unprotected UI sections~\cite{context_hiding_attack}. We show example UI tampering attacks with minimized number of tampering in Figure~\ref{attack:context_hiding}.

\subhead{Security Guarantees}\label{sec:objective}  \cz{}'s objective is to secure requests from a legitimate user to a web server despite a client that has been compromised by malware. It does so by certifying to web servers requests satisfying \textit{interaction integrity} which means that a request 1) came from a real user and 2) was free of tampering by malicious software on the client. \cz{}'s security guarantees only hold if both the user and web server are benign, and it provides no benefits if a legitimate user interacts with a malicious web server (i.e. due to phishing for example), or the user themselves is malicious. \cz{} secures web requests between legitimate users and web servers in 3 phases of user-page interactions: 
\begin{enumerate}[noitemsep,leftmargin=*,nolistsep]
\item \textit{Display}: \cz{} compares the web page that is displayed to the user against a server-provided specification called a VSPEC. For now, we only consider visual content of the web page and do not consider other media such as audio. 
\item \textit{Interaction}: Based on the user's interpretation of the web page, she provides information to the web page. This can include entering data into fields of the page, as well as interacting with other common HTML elements such as drop-down menus, checkboxes, radio buttons, etc. \cz{} infers which inputs the user is providing by tracking standard point of focus (POF) cues, such as a text cursor, and uses this to construct an independent record of the user's inputs.
\item \textit{Submission}: After the user is done entering the information, she submits the contents to construct the request to the web server.  \cz{} then validates that the request is consistent with the user inputs via a validation function in the \Vspec{}. If the function succeeds, the \Vspec{} is included in the request and both are signed by \cz{} certifying the request. The web server can verify that signature, which identifies the \Vspec{} and guarantees that both the display and interaction match the requirements of the \Vspec{}.
\end{enumerate} We summarize the phases of user-page interaction and the attack vectors in Fig \ref{fig:vectors}.

\subhead{Assumptions}
We assume that user request content is constructed from user inputs on a web from and that the user is trustworthy who always provide intended inputs to the web page. We assume the user interacts with the web form in a conventional manner. Specifically, \cz{} depends on the web form employing standard POF cues and that the user pays attention to the values as they enter them into the web page (more on this in Section~\ref{sec:design:intention}). \cz{} does not prescribe an order or manner of data entry, permitting the user to  enter content and then subsequently delete, modify or copy it.  \cz{} also requires the appearance of a web page to be predictable, making excessively unpredictable (i.e. 3rd party ads) or excessively dynamic (i.e. videos) elements unsupported.

\section{\texorpdfstring{\cz{}}{vWitness} Design}\label{sec:design}

In this section, we discuss how the web server and the client should be set up to use \cz{}, and describe the workflow of a typical \cz{} interaction session, followed by a detailed explanation of the \Vspec{} and the interaction validation.

\subsection{Initial Setup}

We discuss the client and web server setup for \cz{}. 

\subhead{Client Device} 
We assume a compromise-free initial setup phase of client devices. During this phase, the user (or a deployment specialist) will 
\begin{enumerate*}[leftmargin=*]
\item Install \cz{}'s hypervisor, which protects \cz{}'s core logic in the dom0. We use Xen's terminology where dom0 refers to the secure virtual machine and domU refers to the guest OS. from malware (including kernel-level malware) in the guest OS (domU).
\item Install a public and private key pair ($K_{pub}$ and $K_{pri}$) and a certificate for $K_{pub}$, $C_{pub}$, certified by a well-known CA. $K_{pri}$ will be used by \cz{} to sign requests and $K_{pub}$ by web servers to validate the requests.

\item Seal $K_{pri}$ to the correct \cz{}'s execution state covering all software components that \cz{} depends on. Successful unsealing of this key ($K_{pri}$) thereafter indicates that the correct \cz{} software stack (i.e \cz{} code and hypervisor) is running, and prevents the exposure of $K_{pri}$ to any principal other than \cz{}. 
In general, sealing requires a measured boot facility and secure element, both of which are commonly available and have been studied in previous work~\cite{xmhf, tan2011tpm, nunes2019vrased, eckel2020secure}. \cz{} assumes the presence of hardware support for sealing.

\item Install an \textit{untrusted} browser extension in domU to facilitate communication between the browser and the trusted \cz{} component.
\end{enumerate*}

\subhead{Web Servers} The server needs to perform a set of one-time modifications to use \cz{}:
\begin{enumerate*}[leftmargin=*]

\item Construct a \Vspec{} for each page requiring \cz{} protection. The \Vspec{} describes a web page's appearance and which elements are used by a user to enter request parameters. In addition, at runtime, unique values, such as a session ID, can be added to a \Vspec{} before transmission to the client to protect against replay. We will describe \Vspec{} content in detail as they are used. 

\item Modify the web page to remove incompatible UI elements if any. We discuss such incompatibilities in \S \ref{sec:limitations}.
\end{enumerate*}

\begin{figure}[t]

\begin{subfigure}{\linewidth}
\centering
 \fbox{\includegraphics[width=0.75\linewidth]{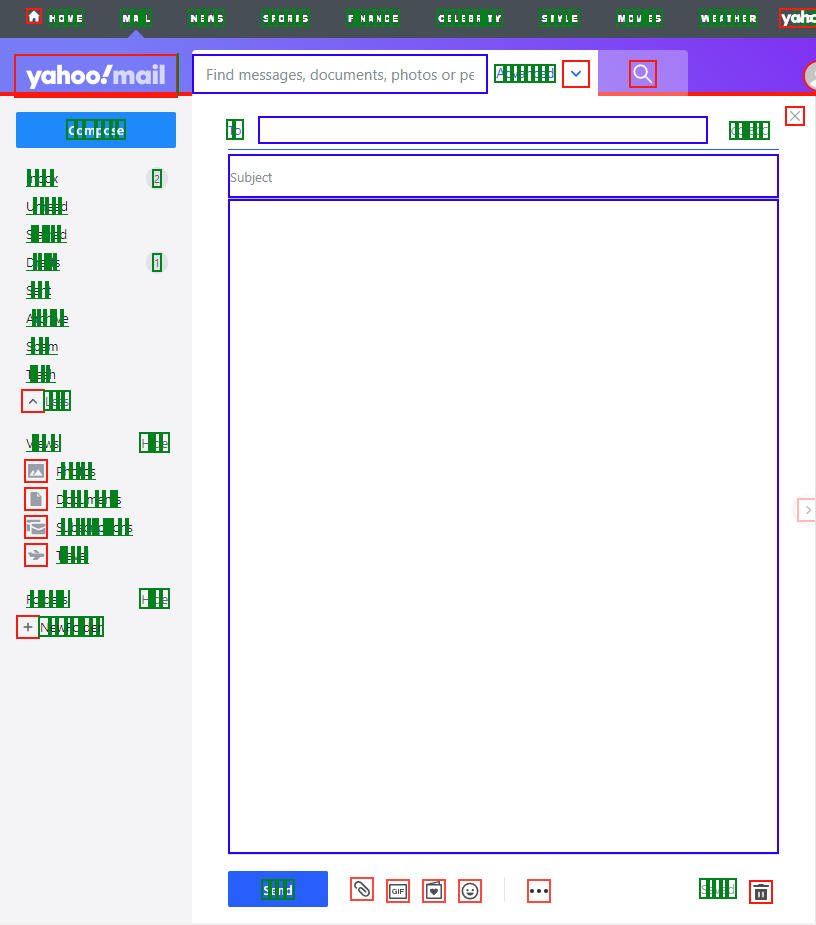}}
 \caption{Color-coded expected appearance.}
 \end{subfigure}
 \vspace{+3mm}

\begin{subfigure}{\linewidth}
\centering
 \fbox{\includegraphics[width=0.8\linewidth, trim=0 1mm 0 2mm,clip]{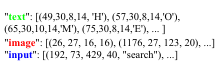}}
 \caption{Elements manifest, element type is color-coded. }
 \end{subfigure}
\caption{\Vspec{} of Yahoo email web client. The expected appearance is cropped to remove personal information.}
  \label{fig:vspec}
  \vspace{-5mm}
\end{figure}

\subsection{Workflow}
\begin{figure}[t]
  \centering
  \includegraphics[width=\linewidth]{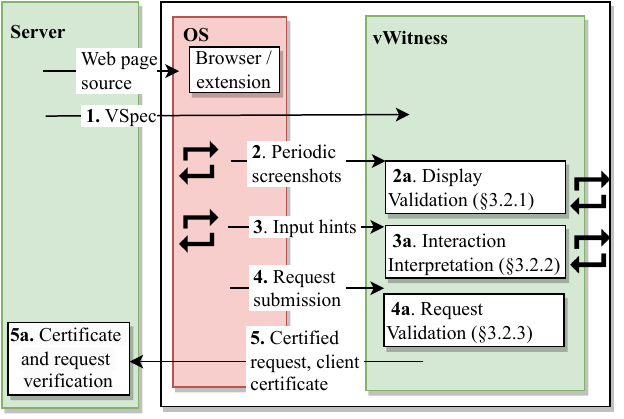}
  \captionof{figure}{\cz{} overview (\S\ref{sec:design}). Green/red indicates trusted/untrusted entities. }
  \label{fig:overview}
  \vspace{-3mm}
\end{figure}

With the client and server set up,  we describe \cz{}'s workflow with references to Fig \ref{fig:overview}.

\subhead{Client-side}  
When a user requests a \cz{}-enabled web page, 
the server replies with the \Vspec{} together with the page source files. The untrusted browser extension forwards the \Vspec{} to \cz{} running on the client (step 1). After \cz{} receives the \Vspec{}, it begins screenshotting by securely accessing the display buffer (step 2).
On every screenshot, \cz{} validates the appearance of the rendered page by comparing to what is expected in the \Vspec{} (step 2a) and, if the user has entered inputs, \cz{} interprets the user-entered input values through input hinting (step 3, we discussed this below). If a user-entered value is to be submitted in the web request according to the \Vspec{}, \cz{} records that value (step 3a). These two steps repeat until the user submits the page. 

When the user submits, the web page logic constructs the web request and uses the browser extension to submit this to \cz{}'s secure component for certification (step 4). \cz{} first validates the content of the request using a server-supplied validation function in the \Vspec{} by executing this function with the inputs it has observed during the user-page interaction (step 4a). If the validation function succeeds, \cz{} creates a signature using $K_{pri}$ over
\begin{enumerate*}
    \item the request content
    \item the \Vspec{} (which contains a session id for freshness)
\end{enumerate*} using the sealed signing key. Signed requests are certified by \cz{} to be constructed with interaction integrity and are returned to the web browser extension, together with the certificate, $C_{pub}$, who forwards them to the web server (step 5).

\subhead{Server-side} 
The web server, upon receiving the certified request and $C_{pub}$, takes the following steps: 
    \begin{enumerate*}
        \item verifies that $C_{pub}$ is signed by a legitimate CA,
        \item verifies the signature on the request using $K_{pub}$ in $C_{pub}$, thus ensuring the integrity of the request content and the included \Vspec{},
        \item verifies that the included \Vspec{} is the \Vspec{} it expects, ensuring that the web page interactions are correct. Note that the \Vspec{} should include a session ID (i.e. a nonce) for freshness.
    \end{enumerate*}   
These three properties allow the server to gain confidence in the interaction integrity of the received request (step 5a).

Note that what is eventually sent out in the request remains unchanged, except the addition of this signature and the corresponding certificate ($C_{pub}$), hence preserving privacy.
While \cz{} has access to the user's UI, and can see intermediate values that the user has entered, as well as other applications on the user's screen, no information about these is leaked to the remote web server. \cz{}'s secure component is intended to be open-sourced and can be scrutinized by the human user.

\subsection{Validating Web Page Interactions}
\label{sec:verification}

\cz{} continuously monitors the Display during the user-page interaction session so that the UI displayed to the user is ensured to be correct throughout the session. In our Xen-based prototype, this is achieved by randomly sampling the domU's virtual frame buffer. The sampling is random so that an attacker cannot predict when \cz{} samples, preventing time-of-check to time-of-use (TOCTOU) attack~\cite{screenshot_sidechannel}.  \cz{} samples with a random delay ranging from 0 to 500ms between every two samples, meaning that on average, \cz{} samples four times a second.  To have a reasonable probability of guessing the sampling times, an attacker would have to change the display much more frequently than 500ms, but previous research has shown that humans may not perceive images that are shown for less than 500ms~\cite{exposure1}. 
We discuss how user-page interactions are validated in the following.

\subsubsection{Display} 
\label{sec:UI} \label{sec:validation_method} \label{sec:region-based_validation}

The page appearance must be untampered. \cz{} does so through a three-step validation: first, \cz{} determines the currently visible portion of the page (view port); second, \cz{} finds all the UI elements within the view port---they need to be validated; finally, \cz{} uses computer vision to validate that each element has been rendered correctly. \cz{} can be configured to validate that all regions without UI elements match a solid background color specified in the \Vspec{}. Non-solid color backgrounds are currently not supported.

To determine the UI elements visible in the browser view port, the \Vspec{} includes the web page's \textit{expected appearance}  (Fig \ref{fig:vspec}), which is a ``long'' screenshot of the page rendered at the client's window width and at the maximum height to accommodate all elements of the page. The client's width is passed to the web server by the untrusted browser extension, whose security implication is discussed in \S \ref{sec:analysis}. \cz{} determines the current view port by matching the sampled frame buffer against the expected appearance and finding the offset from the top of the expected appearance with the best match. %

To figure out the expected UI elements displayed in the view port, \Vspec{} contains an \textit{elements manifest} describing all expected UI elements using type, position and a ground truth. While we detail type and ground truth below, the relevant field for this step is the element position, which is a tuple of (x, y, width, height) illustrating the bounding rectangle of the element. All elements whose bounding rectangle overlaps with the view port are to be validated in the next step.

Finally, \cz{} validates the observed screenshot against the expected elements. 
To do this, \cz{} could naively perform a pixel-by-pixel comparison of the observed element with that in the \Vspec{}, but this would result in many false alarms due to \textit{benign rendering variations} caused by differences in the rendering stack among client devices, which can include different browsers, OSes, device drivers, GPUs, and configuration settings~\cite{rendering_variation1,pixel_perfect}.  
Instead of trying to explicitly identify and handle all differences that could be generated by all such combinations, \cz{} trains convolutional neural network (CNN) models to classify whether the differences between the observed and expected appearances are benign rendering variations.  While machine learning is used in security (e.g. anomaly detection~\cite{juba2015principled} and program vulnerability detection~\cite{7163061}), it is prone to adversarial examples against which we evaluate our approach in \S \ref{eval:ocr}.  

To enable validation, for each expected element, the elements manifest contains its type (text or image) and the ground truth. The ground truth is type-dependent and is how the element should be rendered. For text, it is the actual character; for images, it is the region of the Expected appearance.

\cz{} uses two separate CNN models for text and images as they have different requirements.  For instance, tampering with a few pixels may change the meaning of text (e.g. ``i'' vs ``l''), but semantically identical images may be different pixel-wise (e.g., due to image compression). 
The models take two inputs: the ground truth from the \Vspec{} and the observed element at the same position on the screenshot. 
The models output a boolean prediction indicating whether the observed element matches the expected one. Both models use CNNs for feature extraction and dense layers to compare to the expected rendering.

\subhead{Dynamic Web Elements} 
One challenging aspect of validating web elements is that they can be dynamic. \cz{} categorizes dynamism into three types: scrollable (e.g. slide shows and scrollable list box), dynamically-appearing (e.g. context menu) and dynamically-scaled (e.g. resizeable elements) and develops unique validation methods for each type by tweaking the existing validation method (for static text and images). For brevity, we only discuss the scrollable elements. %
The key insight that enables dynamic element validation is to use a separate \Vspec{}, called nested \Vspec{}, for each dynamic element. This nested \Vspec{} is different from the page's and defines the dynamic element's appearance and how it can be interacted. 

Scrollable elements refer to the UI elements that can scroll independently from the page (e.g. slide show and scrollable listbox). To support its validation, 1) in the page's elements manifest, scrollable elements have two new types to distinguish horizontal and vertical scroll and 2) the expected appearance is all the possible appearances of the scrollable element merged together. The actual validation of scrollable elements is identical to the page's with a change to the view port determination step: view port can either be horizontal or vertical depending on the type in the elements manifest.

\subsubsection{Interaction}
\label{sec:design:intention}

\cz{} must capture user input semantics to validate outgoing requests. While this can be done through intercepting hardware IO inputs and secured device drivers as in previous works~\cite{Fidelius, protection}, we realize that user input is displayed to the user and thus already in the screenshot that \cz{} takes for Display validation. This enables a novel approach where \cz{} extracts user input semantics from the screenshots without the need for secure device drivers. 

We need to make a \textit{reflective validation} assumption, which states that while entering input, human users are generally attentive to whether the inputs are reflected correctly on the display~\cite{gyrus}. For instance, after the user types ``100'' on the keyboard, she will confirm that the value ``100'' is displayed in the intended field, and if not, she will adjust the entered data until it does.  Therefore, regardless of how the user inputs the value, if the user sees it on the display, it is the correct value.

To extract user inputs, a naive approach is to use Optical Character Recognition (OCR) on every input element, but we rejected it as it is computationally expensive and prone to adversarial example attacks~\cite{ocr_ad_blocking}. 
Instead, \cz{} relies on the untrusted browser extension to hint the position and values of user inputs and reuses the CNN-based text validator from the Display phase to ensure the hinted values are correctly displayed. Visual input elements, such as checkboxes and radio boxes, are also encoded as text inputs as their states can be mapped to a well-defined appearance.  To ensure that the browser extension is honest, the elements manifest includes all inputs elements and their positions (in the form of bounding rectangle), \cz{} ensures  that the observed input elements must fall in the bounding rectangle of expected input elements. 

However, extracting inputs from the UI poses two challenges---a subverted client could:
\begin{enumerate*}
\item forge inputs even if a user is not present or 
\item tamper with input elements the user is not paying attention to
\end{enumerate*}. 

\subhead{User Presence} Since hardware I/O observed by the hypervisor cannot be forged by application/OS code, \cz{} assumes user presence if there exist hardware I/O events. Note that \cz{} does not interpret the I/O events but only checks their occurrence from a hardware device, e.g. a keyboard/mouse, enforcing that the timing of these events corresponds to the timing of inputs observed on the UI.

\subhead{User Attention} 
We observe that modern browsers use visual indicators (e.g. focus outline, caret) to indicate which element is currently \textit{in-focus}, which we refer to as the point of focus (POF) and consider the focused element to be where the user's attention is. Assuming users naturally perform reflective validation at the POF, \cz{} only accepts input changes (i.e. insertions, deletions) with POF. \cz{} relies on pixel information to locate POFs, and our prototype currently supports three common POFs: focus outline for elements, input cursors (i.e. caret and vertical line) and multi-character highlight. Because POFs are rendered by the untrusted client, an attacker may tamper with them.  For example, an attacker could forge multiple POFs, perhaps with one more noticeable than others, confusing \cz{} about which POF is actually in use (e.g. the user thinks she is interacting with field A, but \cz{} is validating inputs from field B). \cz{} deals with this by validating the visual consistency of POFs to ensure only one set of POFs is present at any time (further discussed in \S \ref{sec:implementation:client}). Note that tampering with out-of-viewport inputs fields (e.g. due to user scrolling) will be ignored by \cz{}.

\subsubsection{Submission}\label{sec:request_integrity} 
The construction of outgoing requests is server-specific, \cz{} validates requests by executing a server-supplied validation function in the \Vspec{}. 
The validation function takes interpreted inputs and the page request and produces a boolean indicating whether the request validation succeeds. This choice enables flexibility. In the simplest case, the validation function assembles inputs as a JSON object (e.g. for POST requests) and compares that in the page-constructed request as shown in Figure~\ref{fig:vspec}. However, it can implement arbitrary validation logic such as repeating the original construction logic in the web page or validating a set of constraints. This choice is again server-specific.    
The validation function can validate values in the request that are not interpreted by \cz{} such as session IDs to distinguish multiple sessions and nonces for replay prevention; they need to be included into the \Vspec{} by the server. Finally, the \Vspec{} used for validation should be included in the signed request.

\subsection{Limitations} \label{sec:incompatibility}\label{sec:limitations}
\cz{} relies on \Vspecs{} for validation. This means third party elements (e.g. ads iframes) whose content is generated on-the-fly (e.g. from ads providers) are not supported. 
Also, \cz{} can only certify interactions visible on the UI. Invisible interactions (i.e. file uploads) will not be seen by \cz{} and thus cannot be validated unless a checksum is shown on the page.  
While \cz{} can handle some dynamic behavior, the use of nested \Vspec{} makes it computationally too costly to validate excessive dynamicsms (e.g. videos). 
Finally, web pages must use a POF styles (e.g. text highlight color) recognized by \cz{} and use input fields with POFs. Input fields without POF (e.g. button inputs) are not supported.  It is possible for the server to customize POF style and include those in the \Vspec{} for \cz{}'s use, we leave this as a future work.
Since \cz{} is only intended for \emph{security-sensitive pages}, we believe that these limitations will not affect its adoptability.

\section{Implementation}
\label{sec:implementation}

\subsection{Client-side \cz{}} \label{sec:impl:client}

On the client side, \cz{} exposes three new JavaScript APIs to support requests validation:
\begin{enumerate*}
\item \texttt{acquire\_\Vspecs{}}. The extension acquires the client window width through \texttt{Chrome.windows}, forwards it to the web server who returns the \Vspec{} tailored to the client width. The reported width is a virtual pixel value accounting for local settings such as user zooming and OS's default font size. 
\item \texttt{\cz{}\_begin} begins a \cz{}-session where the extension full-screens the page using \texttt{requestFullscreen}, and forwards the \Vspecs{} to \cz{}'s secure component. 
\item \texttt{\cz{}\_end} ends a \cz{}-session. The extension exits full screen using \texttt{exitFullscreen}, and submits a page-constructed request body for validation. %
\end{enumerate*}

\begin{table*}[t]
\centering
\small
\begin{tabular}{l|p{5.8cm}|p{4cm}|r|r}
  \rowcolor{gray!75}     Model        & Inputs                                                                                                                                             & Output                                                                                                            & \# Parameters & \# Training Data \\
\rowcolor{gray!25} Text      &  1. a locally rendered text character (32x32) and
2. the expected character as a string                       &   & 352,097              & 556,512            \\ \cline{1-2} \cline{4-5} %
 \rowcolor{gray!25} Graphics & 1. a (sub-)region of a locally rendered graphics (32x32) and 2. the expected appearance &  \multirow{-3}{4cm}{Whether the local rendering is a benign rendering variation of the expected character/appearance }& 1,761,089            & 620,217           
\end{tabular}
\caption{Validation model details.}
\label{tab:model_details}
\vspace{-5mm}
\end{table*}

\subhead{CNN Models} \cz{} implements two models separately for text and images with model details in Table~\ref{tab:model_details}. Our models are available in both Keras H5 format and Tensorflow Lite and we focus on the English language, but are not limited to it.  We detail the training data collection process. For the text model, we we collect  images of a single character rendered in legitimate rendering environments. We render 94 characters (10 numbers, 52 upper and lower English alphabet characters and 32 symbols) using 231 unique fonts, three styles (Italic, Bold and normal), three renderers (Gecko of Firefox, Blink of Chrome and Webkit of Safari) on two platforms (Windows and MacOS). This yields a core dataset of 8184 pairs of training points (as not all fonts support all characters). We expand the core dataset by 1) enlarging and shifting the characters 2) changing the color intensity and 3) randomly bit-flipping a subset of the pixels yielding a dataset of 278,256 images of legitimate one-character rendering. The purpose of expanding the core datasets is to 1) increase the total number of training data, 2) force the model to learn about the stroke shapes instead of memorizing the position of the characters and 3) augment training data such that random bit-flipped pixels do not affect the prediction \cite{DBLP:journals/corr/HeZRS15}. 
To create a balanced dataset, we introduce data points with a false label by assigning another character to every image. This yields a perfectly balanced training set of 556,512 training data. We optionally trained text models with collapsed expected text (i.e. `s' and `S') to increase the text models' accuracy~\cite{emnist}. 
The training data for the image model is similarly constructed, with the core dataset from a subset of CIFAR-10~\cite{cifar_10_tech_report} and Google's Material icon dataset~\cite{material_icon}. We additionally added false data points with text in the images to ensure that unexpected text in the images will be detected. The total number of training data for the image model is 620,217.

\label{sec:implementation:client}

\subhead{POF Consistency Checks} \cz{} current only supports the following input fields: text boxes  \cz{} extracts POFs through pixel information using OpenCV and enforces the following consistency rules: 1) \textit{Number of instances}. For each POF, \cz{} ensures that the number of POF instances on the context frame to be less than or equal to one. 2) \textit{Same-field Logic}: the focus outline, selection highlight and caret must reside in the same field and 3) \textit{Mutual Exclusivity}: caret and the selection highlight must not be present at the same time.

\subhead{Performance Considerations}
\begin{enumerate*}
\item \textit{Differential Detection}: due to the frequent screenshots, unchanged UI elements do not need to be re-validated. Thus, \cz{} finds the difference between two consecutive screenshots and only validates that region.
\item \textit{Caching}: There are three caches: text, image, and frame. For each, the key is a cryptographic digest of the corresponding display region and the value is the corresponding validation result.
\end{enumerate*}

\subsection{Server-side Scripts}

\cz{} implements the following server-side scripts.

\subhead{Addressing Incompatibilities}  The script handles the limitations in \S \ref{sec:limitations}.  First, we consider all \texttt{iframe}s with a \texttt{src} that points to an external domain as loading nondeterministic content such as ads. Our script removes all such  \texttt{iframe}s. 
Second, our script tries to make user inputs visible where possible, and warns for inputs that might be not visible. The script adds a \texttt{maxlength} field to all textual inputs fields (e.g. \texttt{type=text}, \texttt{textarea}).  
Third, the script searches for POF definitions in all CSS. We warn the developer if the following keywords are present: ``outline'', ``caret'' and ``.focus''.
Finally, the script searches for unsupported HTML elements and warns the developer: files inputs (\texttt{type=file}), drag\&drop inputs (\texttt{ondrop} attribute) and videos ( \texttt{video} tag).

\subhead{Generating \texorpdfstring{\Vspecs{}}{VSPECs}} \label{appendix:ground_truth_script} The script automatically constructs \Vspecs{} by 1) render the web page and 2) annotates HTML elements with the corresponding type for validation using a pre-defined HTML tag-to-validation type mapping. The result of the annotation (Figure \ref{fig:vspec}) is  visually presented to the developer who can manually make adjustments.

\section{Security Analysis} \label{sec:analysis}

\subsection{\cz{}' Security}  
\label{analysis:requests} 
\toconsider{An adversary can seek to manipulate the user's interaction with a web page in any of the Display, Interaction or Submission components.  In addition, the adversary may also seek to mislead \cz{} by tampering with the operation of the untrusted browser extension.    }

\subhead{Server-\cz{} Communication}
\cz{} protects against an adversary who attempt to manipulate communication between the web server and \cz{} as follows:
\begin{enumerate*}[leftmargin=*]
    \item Tampering with web requests: each certified request from \cz{} is signed with $K_{priv}$, which is protected from exposure using cryptographic sealing. 
    \item Tampering with the \Vspec{}: each certified request from \cz{} contains the \Vspec{} that was used to validate the user's interaction with the web page and the user-inputs in the request. The web server should ensure that the \Vspec{} in the request matches the one that it sent with the web page, which prevents an adversary from modifying or replacing the \Vspec{} before \cz{} uses it to validate the user's interaction with the web page.
    \item Replaying an interaction: the \Vspec{} contains a non-repeating session ID to ensure freshness. This prevents an adversary from replaying a previously signed request. Similarly, replacing a \Vspec{} with an old one before \cz{} performs validation will be detected by the web server as the \Vspec{} included in the request will contain a stale session ID.
\end{enumerate*}

\subhead{Display} An attacker cannot show different UIs to \cz{} and the user because \cz{} directly samples from the frame buffer via using Xen's high privilege.  The attacker cannot launch TOCTOU attack: showing the proper display just when \cz{} samples the buffer but shows a tampered UI at other times (when the user perceives) because \cz{} samples at random intervals with a mean of 250ms in every 500ms interval, which is the threshold shown by  previous research~\cite{exposure1}. 

An attacker cannot tamper with or remove text/images because \cz{} verifies the locally-rendered text/images against the ground truth in the \Vspec{} using the text/image verifier. An attacker cannot inject additional text in images because the image model is specifically trained to reject differences between \Vspec{} images and observed images that contain text. \cz{} always verifies all UI elements in the viewport, so everything the user can see is checked by \cz{}. As mentioned earlier, our prototype \cz{} samples frequently enough that an attacker wishing to evade sampling would have to switch the display too frequently for a typical human to perceive the page.

\subhead{Interaction} First, any out-of-viewport input update is ignored by \cz{} (not included in the request) because \cz{} cannot ``see'' it. We first assume an honest browser extension (which hints the input update to \cz{} correctly), dishonest browser extensions are discussed thereafter. An attacker can forge or tamper with any user inputs, but it will be caught by either the user or \cz{} depending on whether the input update has POF. The user will notice input manipulation at the POF because POFs are designed to be seen by a human user, and users already use them for input entry, and thus will try to fix the inputs on the UI if they do not appear as the user expects them to. For input updates without POF, \cz{} will see them as a violation of POF consistency check and ignore the input update. Whenever \cz{} ignores a hinted input update, the input it tracks and the input values tracked by the page will be different and when the submission function executes, it will detect the difference as reflected on the constructed request. 

\subhead{Submission} An attacker may submit the form before the user finishes~\cite{protection}. While \cz{} cannot prevent this, it enforces that the UI must change after the user interaction ends (i.e. with a call to \texttt{\cz{}\_end}). This way, a malicious client cannot silently submit the form without the user noticing it.  A tampered request submitted to \cz{} will be detected by the \Vspec{}'s submission function, while a tampered \Vspec{} will be detected by the web server as \cz{} will include it in the request.

\subhead{Dishonest Browser Extension} \cz{} does not trust any information from the extension: all information is either included in the signature to the server (to be verified by the server, e.g. \Vspec{}) or verified against the information in the \Vspec{} (e.g. hints). To ensure the hints are correct, \cz{} ensures input fields not hinted, do not receive any inputs. We then analyze possible attacks on this interface.

The \texttt{acquire\_\Vspecs{}} call relies on the client window width. An incorrect window width by the browser extension leads to an incompatible \Vspec{} being retrieved and will fail the viewport detection due to mismatch between the observed width and the \Vspec{}'s width.
An attacker making early calls to \texttt{\cz{}\_begin} gains no benefit, as it triggers frame buffer sampling and may fail the display validation depending on what is shown.
The attacker may delay this call until user interaction has begun, and leverage this time difference to show  misleading information.
While \cz{} cannot verify the UI before \texttt{\cz{}\_begin}, it implements the following measures to ensure a clean start of any \cz{}-session to detect delayed \texttt{\cz{}\_begin} calls.
\begin{enumerate*}
\item Clean viewport: the viewport must be at (0, 0)) and
\item Clean input entry: all input fields are empty. \end{enumerate*}
Delaying \texttt{\cz{}\_end} calls does not benefit the attacker as the Display validation continues. Early calls to prematurely end a \cz{}-session will be noticed by the user due to the UI change (e.g., ``submitted'') enforced by \cz{}, who can take corrective actions, e.g., contacting the server.

\subhead{Limitations} \cz{} relies on the secrecy of the private key ($K_{pri}$). If this secrecy is compromised, an adversary can forge \cz{} signatures and certify requests. Mitigating such exposure requires a process for revoking and re-issuing compromised keys. In addition, \cz{} is vulnerable to relaying attacks, meaning a compromised OS can relay an authenticated web session to a malicious user using a correct \cz{} instance. This attack is outside of our current \cz{} threat model, as \cz{} assumes both the user interacting with \cz{} and web server are benign. \cz{} can be extended to handle relaying attacks if user identities are bound to specific \cz{} instances, which might be accomplished with an enrollment procedure similar to that of multi-factor authentication hardware tokens.

\subsection{Adversarial Attacks on CNN Models} \label{eval:ocr}

An attacker can use adversarial examples~\cite{goodfellow2014explaining,adversarial_examples2} to try and fool \cz{}'s CNN-based verifiers. We evaluate \cz{}'s robustness against such attacks by comparing  \cz{} with two reference models: a MNIST classifier~\cite{noauthor_github_nodate} for text and a CIFAR-10 classifier~\cite{noauthor_github_nodate-1} for images.
\toconsider{The reference model g1 is trained against CIFAR-10 since it is not possible to train a classifier on the icon dataset as each icon has a unique label. }
 We believe generic adversarial example defenses (e.g. adversarial training~\cite{szegedy2013intriguing,madry2017towards,DBLP:journals/corr/abs-2103-01946}, gradient masking~\cite{gu2014towards}, detection~\cite{metzen2017detecting}) are complementary to the \cz{}-specific defenses we present here.  However, one such defense, input transformation~\cite{xu2017feature}, is incompatible with \cz{} because it alters the inputs, potentially changing the meaning of a web element.  
Note that as our model takes \emph{unit inputs} (i.e. a single character for text, a 32-by-32 sub-region for image), the accuracy numbers in Table~\ref{tab:t_robustness}  illustrate the worst-case accuracy, as a successful attack will likely need to alter more than one unit input, which exponentially reduces the probability of a successful attack.

\begin{table*}[t]
\centering
\small
\begin{tabular}{p{0.3cm}|p{1.2cm}|r|rrrrr|rrrrrr|p{1cm}}
\rowcolor{gray!50} & &                        & \multicolumn{5}{c|}{L inf}               & \multicolumn{6}{c|}{L2}                                              &\\

\rowcolor{gray!50} \multirow{-2}{*}{\#} & \multirow{-2}{*}{Models}& \multirow{-2}{*}{Clean}                  & FGM    & BIM   & MOM   & FAB    & APGD  & FGM    & BIM    & MOM   & FAB    & APGD  & CW2                     & \multirow{-2}{1.0cm}{Avg attacks} \\

 \multirow{3}{*}{t1} & \multirow{3}{1.2cm}{Reference model}     & \multirow{3}{*}{93.33} & 5.00   & 0.83  & 0.83  & 45.00  & 0.00  & 37.50  & 0.83   & 1.67  & 64.17  & 0.00  & \multirow{3}{*}{6.67}   & \multirow{3}{1.2cm}{13.39 (base)} \\
&&                        & 1.67   & 0.83  & 0.83  & 43.33  & 0.00  & 12.50  & 0.83   & 0.83  & 69.17  & 0.00  &                         &                      \\
&&                        & 0.00   & 0.83  & 0.83  & 43.33  & 0.00  & 7.50   & 0.83   & 0.83  & 68.33  & 0.00  &                         &                    \\

\rowcolor{gray!20}&& & 90.00  & 45.83 & 0.83  & 60.83  & 0.83  & 88.33  & 54.17  & 16.67 & 47.50  & 20.83 &   &  \\
 \rowcolor{gray!20}&&                        & 90.00  & 46.67 & 0.83  & 50.00  & 0.00  & 81.67  & 45.00  & 0.00  & 39.17  & 0.00                       &                  &      \\
\rowcolor{gray!20} \multirow{-3}{*}{t2}& \multirow{-3}{1.2cm}{Base text model}  & \multirow{-3}{*}{98.33}  & 92.5   & 34.17 & 2.50  & 49.17  & 0.00  & 99.17  & 50.00  & 5.00  & 34.17  & 0.00  &\multirow{-3}{*}{7.50}&\multirow{-3}{1.2cm}{37.17 (2.82x)}  \\

\multirow{3}{*}{t3}& \multirow{3}{1.2cm}{Avg one font}       & \multirow{3}{*}{99.36} & 84.87  & 52.42 & 4.62  & 64.29  & 2.83  & 91.29  & 69.00  & 31.42 & 86.92  & 24.63 & \multirow{3}{*}{12.62}  & \multirow{3}{1.2cm}{44.60 (3.38x)} \\
&&                        & 90.37  & 51.88 & 3.71  & 56.50  & 0.00  & 90.38  & 59.92  & 4.96  & 83.29  & 1.37  &                                                & \\
&&                        & 91.62  & 56.04 & 12.58 & 50.21  & 0.00  & 91.71  & 59.46  & 2.33  & 81.96  & 0.42  &                                               & \\
                                    
\rowcolor{gray!20}&&  & 87.92  & 56.83 & 6.25  & 63.42  & 1.33  & 93.58  & 75.50  & 40.00 & 91.42  & 33.42 &   &   \\
\rowcolor{gray!20}&&                        & 89.17  & 49.33 & 5.58  & 53.92  & 0.00  & 91.42  & 63.84  & 7.50  & 87.42  & 2.00  &                         &\\
\rowcolor{gray!20} \multirow{-3}{*}{t4}& \multirow{-3}{1.2cm}{Avg sans serif}&\multirow{-3}{*}{99.50}& 89.17  & 55.00 & 14.17 & 46.58  & 0.00  & 92.33  & 61.92  & 3.50  & 85.67  & 0.50  &\multirow{-3}{*}{16.33}&\multirow{-3}{1.2cm}{47.05 (3.51x)}\\

\multirow{3}{*}{t5}& \multirow{3}{*}{Avg serif}          & \multirow{3}{*}{99.22} & 81.83  & 48.00 & 3.00  & 65.17  & 4.33  & 89.00  & 62.50  & 22.83 & 82.42  & 15.83 & \multirow{3}{*}{8.92}  & \multirow{3}{1.2cm}{43.88 (3.28x)}\\
&&                        & 91.58  & 54.42 & 1.83  & 59.08  & 0.00  & 89.33  & 56.00  & 2.42  & 79.17  & 0.75  &                         &                       \\
&&                        & 94.08  & 57.08 & 11.00 & 53.83  & 0.00  & 91.08  & 57.00  & 1.17  & 78.25  & 0.33  &                         &                      \\
                                    
\rowcolor{gray!20} &&  & 99.59  & 99.59 & 26.66 & 100 & 10.00 & 99.59  & 100 & 76.25 & 100 & 65.41 &  &  \\
\rowcolor{gray!20}& &                       & 98.75  & 98.75 & 13.75 & 100 & 0.41  & 99.17  & 100 & 28.34 & 99.16  & 6.25  &                         &\\
\rowcolor{gray!20} \multirow{-3}{*}{t6}& \multirow{-3}{1.2cm}{High threshold (0.99)} & \multirow{-3}{*}{99.47}& 100 & 89.58 & 12.50 & 98.75  & 0.41  & 100 & 100 & 10.84 & 99.16  & 1.67  &\multirow{-3}{*}{100}&\multirow{-3}{1.2cm}{\textbf{68.86} (5.14x)}

\\ \hline

\rowcolor{gray!50} \multicolumn{15}{c}{Text models above, image models below} \\ \hline

\rowcolor{gray!20} & & & 11.67                   & 8.33                   & 8.33          &0.00&0.00& 22.50                   & 9.17                   & 8.33              &0.00&0.00&   &\\
\rowcolor{gray!20} &&                                            & 9.17& 6.67    & 7.50                   &0.00&0.00& 22.50                   & 9.17                   & 8.33                   &0.00&0.00&                         &\\
\rowcolor{gray!20} \multirow{-3}{*}{g1}& \multirow{-3}{1.2cm}{Reference model} & \multirow{-3}{*}{88.33} & 6.67                   & 3.33                   & 6.67                   &0.00&0.00&18.33&9.17&8.33&0.00&0.00&\multirow{-3}{*}{7.50}& \multirow{-3}{1.2cm}{6.98 (base)}\\

\multirow{3}{*}{g2}&\multirow{3}{1.2cm} {Image model - CIFAR-10}      & \multirow{3}{*}{98.13}                     & 96.67                   & 78.33                   & 77.50                   & 96.67                   & 35.83                    & 93.33                   & 90.83                   & 94.17                   & 97.50                   & 81.67                    & \multirow{3}{*}{12.50}  & \multirow{3}{1.2cm}{75.91 (10.88x)}\\
&&                                            & 98.33                   & 78.33                   & 61.67                   & 95.83                   & 13.33                    & 90.83                   & 81.67                   & 90.00                   & 96.67                   & 53.33                                             &&\\
&&                                            & 100                  & 81.67                   & 59.17                   & 91.67                   & 3.33                     & 95.83                   & 82.50                   & 86.67                   & 96.67                   & 40.83                                            &&\\
\rowcolor{gray!20}&&                     & 93.33                   & 98.33                   & 90.83                   & 97.50                   & 49.17                    & 99.17                   & 99.17                   & 99.17                   & 100                  & 85.83                    &   &\\
\rowcolor{gray!20} &&& 90.00                   & 93.33                   & 79.17                   & 96.67                   & 25.00                    & 95.83                   & 98.33                   & 98.33                   & 99.17                   & 71.67                    &                        &\\
\rowcolor{gray!20} \multirow{-3}{*}{g3}&\multirow{-3}{1.2cm}{Image model - Icon} & \multirow{-3}{*}{99.96} & 94.17                   & 89.17                   & 77.50                   & 94.17                   & 5.83                     & 96.67                   & 98.33                   & 95.00                   & 99.17                   & 54.17                    &\multirow{-3}{*}{28.33}& \multirow{-3}{1.2cm}{83.63 (11.98x)}

\end{tabular}
\caption{The model accuracy (\%) of the text and image models under adversarial examples. The three sub rows in every model correspond to Epsilon=0.1254, 0.2509, 0.5019 for Linf and Epsilon=1, 2, 3 for L2. The accuracy numbers are for unit inputs. }
\label{tab:t_robustness}
\vspace{-3mm}
\end{table*}

\subhead{Attack Setup} We measure the models' accuracy (as a way to measure robustness) under the following attacks: fast gradient method (FGM)~\cite{goodfellow2014explaining}, basic iterative method (BIM)~\cite{DBLP:journals/corr/KurakinGB16}, momentum (MOM)~\cite{DBLP:journals/corr/abs-1710-06081}, CW~\cite{cw_adv_exp} with L2 distance (CW2), AutoPGD with cross entropy loss (APGD)~\cite{croce2020reliable} and Fast Adaptive Boundary (FAB)~\cite{DBLP:journals/corr/abs-1907-02044}. We use Cleverhans v4.0~\cite{papernot2018cleverhans} for the first four and AutoAttack~\cite{croce2020reliable} for the rest. All attacks are \textit{targeted} attacks fooling the model into producing the opposite label. We launch attacks in three epsilon values (i.e. maximum allowed perturbations) measured using two distances (the three sub rows of each model in Table~\ref{tab:t_robustness}). These epsilon values are picked to cover the range of successful attacks~\cite{cw_adv_exp} to best determine the models' robustness. 
We tune hyper-parameters to give the highest attack success rate. We use the average accuracy of all attacks to measure a model's robustness.
\toconsider{: epsilon(E)=0.1254, 0.2509, 0.5019 for Linf norm (which translates to a maximum tampering of 32, 64, 128 out of 255) and epsilon(E)=1, 2, 3 for L2 norm. For the image model, we separately generate attacks for CIFAR-10~\cite{cifar_10_tech_report} and Material icon~\cite{material_icon}.  }
We round all generated attacks to the nearest pixel value between 0 and 255 to make them valid images. 

\subhead{Base Robustness} We tabulate the accuracy of the base \cz{} models (t2, g2 and g3) and the reference models (t1 and g1) in Table~\ref{tab:t_robustness}. We find that \cz{}'s base text and image models are more robust by a factor of 2.82 and 11.98 respectively. We believe \cz{}'s high base robustness is due to the following 
\begin{enumerate*}
\item \cz{}'s models perform a simpler task: While the reference models perform multi-label classification, \cz{} performs binary classification between the Display and the ground truth from \Vspecs{}. 
\item The ground truth in \Vspecs{} further reduces the attack surface as only one targeted attack is applicable. The attack must cause the model to misclassify, from a false prediction into a true prediction (e.g. the attacker wants to change the word ``yes'' into ``no'' on display, but still have it classified as a ``yes'' so it matches the \Vspec{}). The opposite attack (from true into false) is not applicable to \cz{}. 
\end{enumerate*}
Since the image models are already 10.88 and 11.98 times more robust, in the rest of the section, we focus on defenses specific to \cz{}'s text models.

\noindent \textit{Specialized single font models} can be adopted if web servers are willing to use a pre-selected font for all text on the page. This restricts the input space, allowing a smaller adversarial space, which in turn, increases the model robustness. Note that this is different from input transformation~\cite{xu2017feature}. We trained 20 text verifier models, each specialized to a single font and report the robustness in Table~\ref{tab:t_robustness} t3. Single font increases the model robustness by a factor of 3.38. In this experiment, since we aim to explore the robustness change due to input space change, we did not change the model capacity (i.e. reducing layers or the number of training parameters)---so there is potential for further robustness due to  distillation~\cite{papernot2016distillation}.

\subhead{Font Characteristics} We further examine whether a font's characteristics have an impact on a model's robustness. We evaluated fonts with varying characteristics such as font weight, width and type~\cite{sadko_guide_2021,gelderman1999short}. The only significant factor we found is serif or sans serif. We trained 10 serif and 10 sans-serif single-font models and report the average accuracy in rows t4 and t5 of Table~\ref{tab:t_robustness}. The robustness increases by a factor of 3.51 and 3.28 for sans serif and serif font types respectively.

\begin{table}[t]
\centering
\small
\begin{tabular}{c|cccccc}
\rowcolor{gray!55}      Attacks               & \multicolumn{3}{c}{MOM}    & \multicolumn{3}{c}{APGD}                                                 \\ \cline{2-7}
\rowcolor{gray!55} Epsilon        & 1 & 2 & 3 &  1 & 2 & 3   \\ 

 \cellcolor{gray!55}
 
      Original 

  & \raisebox{-3mm}{\includegraphics[width=.075\linewidth,trim=0 3mm 0 0, clip]{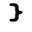}  }
&  \raisebox{-3mm}{\includegraphics[width=.075\linewidth,trim=0 3mm 0 0, clip]{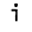}}
& \raisebox{-3mm}{\includegraphics[width=.075\linewidth,trim=0 3mm 0 0, clip]{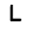} }
& \raisebox{-3mm}{\includegraphics[width=.075\linewidth,trim=0 3mm 0 0, clip]{pic/adv_examples/orig_1_7_p.png}}
& \raisebox{-3mm}{\includegraphics[width=.075\linewidth,trim=0 3mm 0 0, clip]{pic/adv_examples/orig_2_9_-.png}}
& \raisebox{-3mm}{\includegraphics[width=.075\linewidth,trim=0 3mm 0 0, clip]{pic/adv_examples/orig_5_t.png}}
 \\ \cline{2-7}
\cellcolor{gray!55} Target & p & -  & t  & \{  & e & A
 \\ \cline{2-7}
 \cellcolor{gray!55} 
 
Attacks

& {\raisebox{-4mm}{\includegraphics[width=.075\linewidth,trim=0 2mm 0 0, clip]{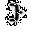}  }}
&  \raisebox{-4mm}{\includegraphics[width=.075\linewidth,trim=0 2mm 0 0, clip]{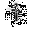}}
& \raisebox{-4mm}{\includegraphics[width=.075\linewidth,trim=0 2mm 0 0, clip]{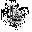} }
& \raisebox{-4mm}{\includegraphics[width=.075\linewidth,trim=0 2mm 0 0, clip]{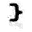}}
& \raisebox{-4mm}{\includegraphics[width=.075\linewidth,trim=0 2mm 0 0, clip]{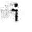}}
& \raisebox{-4mm}{\includegraphics[width=.075\linewidth,trim=0 2mm 0 0, clip]{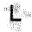}}
 \\

\end{tabular}
\caption{Successful adversarial examples generated for \cz{}'s t6 model in Table~\ref{tab:t_robustness} using L2 norm.}
\label{tab:adv_examples}
\vspace{-4mm}
\end{table}

\noindent \textit{High Detection Threshold} increases the model robustness by a factor of 5.14. A high detection threshold forces the attacker to produce more convincing attack examples but may reduce the model accuracy (as the model needs to produce more convincing predictions). We trained three single-font sans serif models with three thresholds: 0.75, 0.9 and 0.99. We show the result of the 0.99 threshold in Table~\ref{tab:t_robustness} t6.  The threshold at which certain amount of adversarial perturbation becomes user-noticeable is highly subjective. However, prior work shows that adversarial examples cannot fool a time-unlimited human~\cite{elsayed2018adversarial}. What is more, we believe perturbations added to typeset text are more noticeable than perturbations added to hand-written text. We put a random subset of successful adversarial examples in Fig~\ref{tab:adv_examples}. FGM, BIM and MOM do not minimize perturbation, so their perturbation seems more ``randomly'' spread out and, thus, is (in our opinion) easily perceivable and has limited effectiveness. In practice, an attacker will likely need to change multiple characters to alter the meaning of the text, making an adversarial example attack even less likely to succeed against \cz{}.

\subsection{Trusted Computing Base (TCB) Size}  \label{analysis:tcb}

\begin{figure}[t]
\small
\rowcolors{1}{white}{gray!20}
\centering
\begin{tabular}{cr}
\rowcolor{gray!50} Description & Lines of Code \\
\cz{}  &  1,128      \\  
WolfCrypt    &  2,801 \\ 
OpenCV     & 177,396  \\ 
 Tensorflow Lite &14,580 \\
  Xen & 555,160 \\\hline

 Chromium & 25,163,547 \\ 
 Firefox & 20,928,358 \\
\end{tabular}
\captionof{table}{The TCB size of \cz{} prototype's main components.}
\label{table:tcb}
\vspace{-1mm}
\end{figure}
The core logic of \cz{} itself was implemented in just over 1K lines of code (LOC), making it small and auditable, but it depends on several external components, such as OpenCV for processing raw images from the frame buffer, Tensorflow for computer vision and CNN models, and Xen \& QEMU for isolation.  We show \cz{}'s LOC breakdown in Table~\ref{table:tcb}. 
For comparison, we also list the code sizes of the Chromium and Firefox browsers.  Note that our \cz{} prototype was designed for ease of implementation as opposed to TCB minimization, and thus uses commercially off-the-shelf components, which make up the vast majority of its TCB.  Many of these components can be substituted with smaller-TCB alternatives (e.g. XMHF~\cite{xmhf}, debloated libraries ~\cite{quach2018debloating}). 
We leave this as future work. \cz{}, at its current stage, already has a much smaller TCB than fully-functional commodity browsers.

\section{Evaluation} \label{sec:eval}

\begin{table}[t]
\centering
\begin{tabular}{llllll}
\rowcolor{gray!75}  & & \multicolumn{2}{l}{Avg Invocations} & \multicolumn{2}{l}{Total Invocations} \\
\rowcolor{gray!75} \multirow{-2}{*}{Dataset} &   \multirow{-2}{*}{\# datapoints}                               & T                 & G               & T                 & G                 \\
Clickbench               & 39                             & NA                & 880             & NA                & 34320             \\
\rowcolor{gray!25} Jotform                  & 100                            & 464.14            & 17.27           & 46414             & 1727             
\end{tabular}
\caption{Complexity of the evaluation datasets. T/G refers to the text/graphics model.}
\label{tab:invocations}
\vspace{-3mm}
\end{table}

\toconsider{We evaluate three aspects of \cz{}.  First, we evaluate \cz{}'s accuracy at detecting UI manipulation attacks using three datasets of existing UI attacks, possible attacks and a benign dataset of web forms.  Second, we evaluate the delay \cz{}'s certification introduces to the submission of a web request.  Finally, we evaluate \cz{}'s compatibility against a corpus of commonly used web pages.  }

We run empirically evaluation on a machine with an Intel i7-7700 and 16GB RAM with two setups: CPU-only and with a GPU. The CPU is always configured as 7 logical cores for the domU and 1 logical core for the dom0 (where \cz{} runs) and 1 GB of RAM to best mimic a production setup. The GPU setup uses a Nvidia 1060 GPU. The GPU is not a requirement for \cz{}'s deployment, but part of a typical workstation and it accelerates \cz{}'s vision-based validators. We use the text model trained with collapsed labels. %

\subsection{Accuracy}
\cz{}'s accuracy is evaluated on one dataset of attacks and one dataset of benign pages to measure the rate of true positives and true negatives. 

\begin{figure}[t]
\rowcolors{2}{gray!25}{white}
\small
\centering
\begin{tabular}{cccc}
\rowcolor{gray!70}
Dataset    & TP/TN   & FP/FN & Accuracy \\
 Clickbench     &  39     & 1  & 97.5\%   \\
Jotforms      & 100 & 0      & 100\%    
\end{tabular}  
\captionof{table}{Accuracy of \cz{}'s output validator (a single display frame). }
\label{tab:accu_perf}
\vspace{-3mm}
\end{figure}
\subhead{Clickbench}~\cite{clickshield} is a corpus of $1080\times1920$ screenshots of simulated clickjacking attacks on Android (Clickjacking falls under our attack model). We acknowledge that while Clickbench being originally targeted at Android UIs is not ideal for evaluating \cz{}, which is designed for web UIs, it is the only available dataset we are aware of that contains malicious UI tampering. We cannot construct \Vspecs{} on Clickbench's screenshots (no HTML), thus, we create a pseudo-\Vspec{} classifying the whole screenshot as a single image invoking \cz{}'s image model only. 
As not all samples are applicable to our evaluation due to different definitions of attacks, we have 40 pairs of Android UIs. The true positive (TP) and false negative (FN, i.e. malicious tampering not detected by \cz{}) rates are reported in Table~\ref{tab:accu_perf}.
There was a single false negative on an attack tampering with text in an image.  We invoked the text model and it was able to flag the tampering. This suggests that our image model alone can handle most Clickjacking attacks.

\subhead{Jotforms} We evaluate \cz{}'s ability to handle benign rendering variations with a set of forms derived from the Jotforms (\url{jotform.com}).  We chose Jotforms as they provide representative samples of many common forms, which are used on over 10 million websites. We randomly select 100 forms  and render each form on several different rendering stacks.  We execute the \cz{}'s scripts to construct the \Vspecs{} in one rendering stack and we render the same pages in other (and different) rendering stacks, which we provide to \cz{} as captured frame buffers. We tabulate the result in Table~\ref{tab:accu_perf} that shows high accuracy of \cz{}'s vision-based validators.

\subsection{Request Delay} \label{sec:perf_eval}

Since \cz{}'s validation is concurrent to the user session, a delay ($L$) is added to the final request, for which we aim to empirically evaluate in this section. This delay can be modelled as $L \!= T(init) \!+ \sum T(frame_i) \!+ T(request) \!- T(session)$ where $T(init)$ is the time to initialize \cz{} (e.g. transmitting \Vspec{}), $T(frame_i)$ is the time to validate the $ith$ display frame and  $T(request)$ is the time to validate request (i.e. executing the validation function and signing) and $T_{session}$ is the length of user's session. We further distinguish the time to validate first display frame $T(frame_0)$ from subsequent frames $T(frame_i)$ for $i > 0$ due to caching. 

\begin{table}[t]
\rowcolors{2}{gray!25}{white}
\small
\centering
\begin{tabular}{p{1.5cm}|c|cccc}
\rowcolor{gray!70}
Setup & Dataset    & Mean   & Max  & Min & Stdev\\

\rowcolor{white} & Clickbench   & 3.29         & 3.63    & 1.68     & 0.65  \\
 \rowcolor{white} \multirow{-2}{1.5cm}{CPU} & Jotforms & 1.17& 6.83   &0.34&0.85\\
 \rowcolor{gray!25}& Clickbench    & 0.73      & 1.67    & 0.44      & 0.18     \\
\rowcolor{gray!25}  \multirow{-2}{1.5cm}{GPU} & Jotforms     & 0.88         & 6.75    & 0.33     & 0.33      \\ \hline
\end{tabular}
\captionof{table}{\cz{}'s performance numbers, in seconds (s), of first display frame ($T(frame_0)$). CPU setup uses a single logical core. }
\label{tab:perf}
\vspace{-1mm}
\end{table}

\begin{figure}[t]
  \centering
  \includegraphics[width=0.8\linewidth]{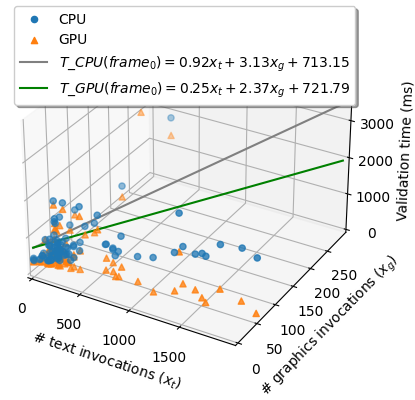}
  \captionof{figure}{First display frame validation time (in ms) of Jotforms. }
  \label{fig:3d}
  \vspace{-5mm}
\end{figure}

\subhead{First Display Frame} $T(frame_0)$ is dependent on the number of model invocations---more complex pages have a higher number of model invocations and thus a larger $T(frame_0)$. We first report the statistics on model invocations in Table~\ref{tab:invocations}.
We report $T(frame_0)$ in Table~\ref{tab:perf}. 
GPU speedup is more significant for Clickbench than Jotforms because the former 1) invokes only the graphics model and 2) have many more invocations than the latter as we treat the whole UI as a single image. 
To know the exact relationship between $T(frame_0)$ and text and graphic model invocations ($x_t$ and $x_g$ respectively), we plot evaluation numbers of Jotforms in Figure~\ref{fig:3d} and fit two regression lines whose coefficients enable the prediction of $T(frame_0)$ of arbitrary complex page. 
As the figure shows, it is more expensive to invoke  the graphic model as it takes two graphics as input and has to do two feature extractions.

\begin{figure}[t]
  \vspace{-14pt}
  \centering
  \includegraphics[width=0.85\linewidth]{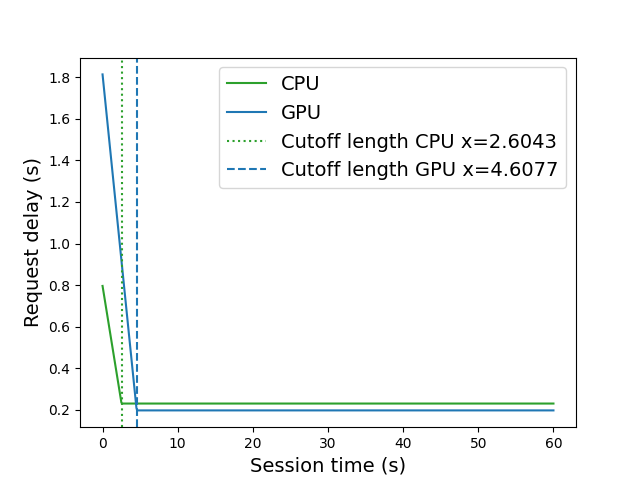}
  \captionof{figure}{Request delay (s) vs session time (s).}
  \label{fig:session_latency}
  \vspace{-1mm}
\end{figure}

\subhead{Subsequent Display Frames} $T(frame_i)$ is page-dependent, and due to caching, it is also interaction-dependent. To simulate  interactions, we recorded our own interactions of filling out a form, a typical use case for \cz{}. 
We report $T(frame_i)$ in the middle four columns in Table~\ref{tab:perf_subsequent} together with measured $T(init)\!+T(frame_0)$ and $T(request)$.

Since $T(frame_i)$ is much less than 1 second, this means that, with a long enough $T_{session}$, there will be a session time threshold $thresh_i$ where \cz{} has finished validating all $frame_{i}$ for $i\!<thresh_i$. We refer such $thresh_i$ as cutoff session length, and session that lasts longer than this time should only incur a delay of $T(request)$. For session length longer than cutoff session length, the request delay $L$ will be $T(frame_last)+T(request)$, which is $0.194\!+\!0.036\!=\!0.230$ seconds for the CPU setup and $0.161\!+\!0.036\!=\!0.197$ seconds for the GPU setup in Table~\ref{tab:perf_subsequent}, 197ms and 230ms are roughly 9\% of the average load time of a web page of 2.2s at the time of this paper~\cite{loading_speed}. To determine the cutoff session length, we plot request delay ($L$) as a function of varying $T_{session}$ in Figure~\ref{fig:session_latency}. The cutoff session lengths are 2.6 and 4.6 seconds for the CPU and GPU setup respectively.

\begin{table}[t]
\centering
\small
\begin{tabular}{c|p{1.2cm}|rrrr|r}
\rowcolor{gray!50}
 && \multicolumn{4}{c}{\small Subsequent frames} &  \\ \cline{3-6} 
\rowcolor{gray!50}
\multirow{-2}{0.5cm}{Setup}& \multirow{-2}{0.9cm}{\small Init+First frame}& Mean     & Max     & Min    & Stdev   &\multirow{-2}{*}{ \begin{tabular}[c]{@{}l@{}}Validation \\function\end{tabular}}  \\ 
CPU &0.760 &0.194 &0.434 &0.088 &0.067 & 0.036 \\
GPU &1.778 &0.161 &0.315 &0.070 &0.055& 0.036 

\end{tabular}
\caption{End-to-end performance numbers, in seconds (s).}
\label{tab:perf_subsequent}
\vspace{-3mm}
\end{table}

\subsection{Compatibility}  \label{analysis:compatibility} \label{sec:tradeoff}

 \begin{table}[t]
\small
\centering
\rowcolors{1}{white}{gray!20}
\begin{tabular}{lr}
\rowcolor{gray!50} System & \# of Compatible Pages \\
Fidelius~\cite{Fidelius}  & 20 (0.77\%)        \\
ProtectiON~\cite{protection}                                      & 196 (7.58\%)        \\
\cz{}                                            & 2,255 (87.23\%) \\
\end{tabular} 
\caption{\cz{}'s compatibility compared to limited-functionality, high-assurance works.}
\label{tab:tradeoff}
\vspace{-5mm}
\end{table}

We measure \cz{}'s ability to support common web forms by constructing a dataset of representative web forms.  
We had initially sought to build this dataset by crawling the Internet for forms, but found that this did not produce a high-quality dataset because:
\begin{enumerate*}
\item security-sensitive forms that \cz{} is meant to protect are often only accessible with authenticated user sessions that our crawler could not access and 
\item publicly accessible forms (with \texttt{form} tags), in our experience, tend to be dominated by simplistic ``contact us'' forms that lack variety and complexity. 
\end{enumerate*}
We thus instead crawled \textit{all} 2476 forms from JotForm and \textit{all} 109 templates from WPForms(\url{wpforms.com}) for Word Press totaling 2585 forms. We did not remove any page from the dataset.

We compare \cz{}'s compatibility with proposed limited-functionality high-assurance web clients \cite{Fidelius,protection}.  These proposals implement web clients that render a subset of HTML elements, but do so with very small TCB.  We find that the smaller TCB is costly: none of them can support any of the pages from our dataset of representative forms, indicating that real web forms still contain many complex web elements that previous work does not account for.  To enable comparison, we relax our requirements and count the number of forms where at least $90\%$ of the elements are supported, to be ``compatible''.  As shown in Table~\ref{tab:tradeoff} previous work can support no more than $8\%$ of real web forms while \cz{} can support almost all forms, with this $90\%$-support compatibility threshold.

\section{Related Works}

\noindent \textbf{Clickjacking} tricks a user into performing unintended actions through deceptive user interface~\cite{clickhijacking,clickshield,context_hiding_attack}. There are current defenses through in-browser enforcement such as framebusting~\cite{frame_bustering1} which can be disabled or bypassed (i.e. directly writing to frame buffer) by privileged malware such as Scranos. \cz{} defends against clickjacking by validating the UI presented to the user. 

\subhead{Request Forgery} 
Botnets forge communications with the service without users' awareness. While there exist defenses attempting to identify such traffic~\cite{zhao2009botgraph} from the service side, well-adopted client-side defenses, such as Captcha~\cite{von2008recaptcha}, fail to bind requests (i.e. bot-generated) to human users, hence Captcha farms. \cz{} binds outgoing requests to a \cz{}-session described in a \Vspec{}, which guarantees association with real human user interaction, preventing bot-generated traffic. 

\noindent \textbf{Cross-site Request Forgery (CSRF)}~\cite{khodayari2021jaw} forges requests to the service by hijacking the user's authenticated session. 
Service-side defenses require requests to provide proof of user session abstracted by tokens~\cite{zeller2008cross} or first-party-only cookies (SameSite)~\cite{SameSite72:online}. These defenses achieve security through obscurity --- a strong client-side malware can access and present these session abstractions to the service bypassing the defense.
An adversary hijacking a \cz{} session can forge requests in the background, but only user-intended requests will receive certification due to display validation and user presence.

\noindent \textbf{Malicious Scripts in Page Context} can do harm such as stealing cookies, forging requests, and tampering with requests~\cite{protection}. \cz{} does not prevent malicious script execution on the client side, it ensures that malicious requests will not receive certification, as they lack a display and user presence.

\subhead{Certifying Requests} 
A common solution adopted by the industry is confirmation through an out-of-band channel. For instance, web services send notification emails to the user after critical actions~\cite{WhydoIge64:online, TradingA12:online}, the response (or its absence) to which is treated as user confirmation. Confirmations have the following drawbacks 1) reactive measure. Confirmation is sent after the service has executed the critical action, meaning that attacks cannot be prevented in the first place; 2) lack of trust. The out-of-band channel may not always be secure and 3) an extra burden on the user side. As a result, the application of out-of-band confirmation is limited. 

Client-side trusted execution defenses enabling secure display~\cite{trusted_display}, secure I/O~\cite{trusted_path} and secure JavaScript execution~\cite{goltzsche2017trustjs, Shuang} are the fundamentals to secure interaction integrity. Below, we will describe academic works in the area. 
\begin{enumerate*}
    \item \emph{Heuristic-based.} Earlier works~\cite{binder,Not-a-bot} leverage the time since the last hardware I/O as a heuristic to certify requests, as hardware I/O can only be generated from the physical human user. Requests without recent hardware I/O are said to be non-user-generated. These works neither verify what is shown to the user nor link hardware I/O to request semantics---an attacker can evade detection by only forging requests when there are I/O activities.
    \item \emph{Linking on-screen inputs to requests.} Gyrus~\cite{gyrus} uses a secure VM to acquire the on-screen user inputs (e.g. email content on the screen) to match with outgoing requests (e.g. email content in the request). Despite that a Gyrus user may see a completely tampered page, Gyrus uses per-field security indicators to indicate whether the on-screen input values have changed since the last time the user has viewed it, which imposes a high cognitive load on users.
    \item \emph{Secure user-page interaction.} VButton~\cite{VButton}, Fidelius~\cite{Fidelius} and ProtectiON~\cite{protection} realize that the display/interaction/submission must all be secured to protect requests. These TEE-based works implement a minimal renderer/device drivers/JS engine inside a secure container (e.g. SGX, TrustZone). These works face a dilemma between functionality support and a low-trusted computing base (TCB). By favoring a smaller TCB, they are extremely limited in terms of  UI elements/input devices/JS features support in the secure area of the page. As a side effect, these works have to allow the coexistence of secured and unsecured UI elements, and thus, are open to UI redressing attacks~\cite{context_hiding_attack} tampering with all the unsecured web elements.
    \item \emph{Secure confirmation} is similar to out-of-band confirmation except the confirmation is shown on the same client securely through a secure execution environment such as ARM TrustZone~\cite{VButton}, Intel TXT~\cite{utp}. These confirmation dialogues display at a fixed time (i.e. after the transaction, before the request is sent). An attacker can exploit this timing by showing a deceptive context before and after the secure dialogue to lure the user into falsely confirming the transaction~\cite{aware}. 
    \cz{} enforces display integrity throughout a \cz{} session. 
\end{enumerate*}

\section{Conclusion} \label{sec:future}

We shift away from directly implementing trusted web clients in hardware-supported TEEs, which can lead to limited functionality and incomplete UI protection, to only relying on passively ``witnessing'' user interactions with the web page, and validating them against a \Vspec{}. We have tested this approach using the \cz{} prototype, which employs CNN-based validators to process raw screenshots of the browser view and certifies the outgoing requests if the validation succeeds based on the provided \Vspec{}. Our evaluation shows that \cz{} is effective in detecting UI tampering, offers significantly better compatibility in terms of supporting real-life web pages, and only introduces negligible latency. Despite the additional effort involved, with the automated \Vspec{} generation, we believe \cz{} will provide an appealing option to security-critical websites.

\bibliographystyle{IEEEtran}

\bibliography{main}

\begin{thebibliography}{10}
\providecommand{\url}[1]{#1}
\csname url@samestyle\endcsname
\providecommand{\newblock}{\relax}
\providecommand{\bibinfo}[2]{#2}
\providecommand{\BIBentrySTDinterwordspacing}{\spaceskip=0pt\relax}
\providecommand{\BIBentryALTinterwordstretchfactor}{4}
\providecommand{\BIBentryALTinterwordspacing}{\spaceskip=\fontdimen2\font plus
\BIBentryALTinterwordstretchfactor\fontdimen3\font minus
  \fontdimen4\font\relax}
\providecommand{\BIBforeignlanguage}[2]{{%
\expandafter\ifx\csname l@#1\endcsname\relax
\typeout{** WARNING: IEEEtran.bst: No hyphenation pattern has been}%
\typeout{** loaded for the language `#1'. Using the pattern for}%
\typeout{** the default language instead.}%
\else
\language=\csname l@#1\endcsname
\fi
#2}}
\providecommand{\BIBdecl}{\relax}
\BIBdecl

\bibitem{dhamija2000deja}
R.~Dhamija and A.~Perrig, ``Deja {Vu--A} user study: Using images for
  authentication,'' in \emph{9th USENIX Security Symposium (USENIX Security
  00)}.\hskip 1em plus 0.5em minus 0.4em\relax Denver, CO: USENIX Association,
  Aug. 2000.

\bibitem{9152694}
S.~Ghorbani~Lyastani, M.~Schilling, M.~Neumayr, M.~Backes, and S.~Bugiel, ``Is
  {FIDO2} the kingslayer of user authentication? a comparative usability study
  of fido2 passwordless authentication,'' in \emph{IEEE Symposium on Security
  and Privacy (SP)}, 2020, pp. 268--285.

\bibitem{ross2005stronger}
B.~Ross, C.~Jackson, N.~Miyake, D.~Boneh, and J.~C. Mitchell, ``Stronger
  password authentication using browser extensions,'' in \emph{14th USENIX
  Security Symposium (USENIX Security 05)}.\hskip 1em plus 0.5em minus
  0.4em\relax Baltimore, MD: USENIX Association, Jul. 2005.

\bibitem{fingerprinting_authentication}
N.~Andriamilanto, T.~Allard, G.~Le~Guelvouit, and A.~Garel, ``A large-scale
  empirical analysis of browser fingerprints properties for web
  authentication,'' \emph{ACM Transactions on the Web (TWEB)}, vol.~16, no.~1,
  pp. 1--62, 2021.

\bibitem{VButton}
W.~Li, S.~Luo, Z.~Sun, Y.~Xia, L.~Lu, H.~Chen, B.~Zang, and H.~Guan, ``Vbutton:
  Practical attestation of user-driven operations in mobile apps,'' in
  \emph{Proceedings of the 16th Annual International Conference on Mobile
  Systems, Applications, and Services}, ser. MobiSys '18.\hskip 1em plus 0.5em
  minus 0.4em\relax New York, NY, USA: ACM, 2018, pp. 28--40.

\bibitem{protection}
A.~Dhar, E.~Ulqinaku, K.~Kostiainen, and S.~Capkun, ``{ProtectIOn}:
  Root-of-trust for io in compromised platforms,'' in \emph{Proceedings of the
  2020 Network and Distributed System Security (NDSS) Symposium}, vol. 2020,
  2020, p. 869.

\bibitem{clickhijacking}
L.-S. Huang, A.~Moshchuk, H.~J. Wang, S.~Schecter, and C.~Jackson,
  ``Clickjacking: Attacks and defenses,'' in \emph{21st USENIX Security
  Symposium (USENIX Security 12)}.\hskip 1em plus 0.5em minus 0.4em\relax
  Bellevue, WA: USENIX Association, Aug. 2012, pp. 413--428.

\bibitem{khodayari2021jaw}
S.~Khodayari and G.~Pellegrino, ``{JAW}: Studying client-side {CSRF} with
  hybrid property graphs and declarative traversals,'' in \emph{30th USENIX
  Security Symposium (USENIX Security 21)}, 2021, pp. 2525--2542.

\bibitem{XFrameOp32:online}
MDN, ``X-frame-options - http $|$ {MDN},'' aug 2022,
  \url{https://developer.mozilla.org/en-US/docs/Web/HTTP/Headers/X-Frame-Options}
  [Accessed Nov 26 2022].

\bibitem{CrossOri34:online}
------, ``Cross-origin resource sharing ({CORS}) - {HTTP} | mdn,''
  \url{https://developer.mozilla.org/en-US/docs/Web/HTTP/CORS} [Accessed Aug 29
  2022], 2022.

\bibitem{ContentS85:online}
------, ``Content security policy ({CSP}) - {HTTP} $|$ {MDN},'' 2022,
  \url{https://developer.mozilla.org/en-US/docs/Web/HTTP/CSP} [Accessed Nov 26
  2022].

\bibitem{Subresou12:online}
------, ``Subresource integrity - web security $|$ {MDN},''
  \url{https://developer.mozilla.org/en-US/docs/Web/Security/Subresource_Integrity}
  [Accessed Nov 26 2022], aug 2022.

\bibitem{agarwal2021first}
S.~Agarwal and B.~Stock, ``First, do no harm: Studying the manipulation of
  security headers in browser extensions,'' in \emph{Workshop on Measurements,
  Attacks, and Defenses for the Web (MADWeb)}, 2021.

\bibitem{BREXKILI31:online}
T.~MICRO, ``Brex\_kilim.ll - threat encyclopedia,'' 2014,
  \url{https://www.trendmicro.com/vinfo/us/threat-encyclopedia/malware/BREX_KILIM.LL/}
  [Accessed Nov 26 2022].

\bibitem{scranos}
TechCrunch, ``\BIBforeignlanguage{en-US}{Scranos, a new rootkit malware, steals
  passwords and pushes {YouTube} clicks},''
  \url{https://social.techcrunch.com/2019/04/16/scranos-rootkit-passwords-payments/}
  [Accessed Nov 26 2022].

\bibitem{Not-a-bot}
R.~Gummadi, H.~Balakrishnan, P.~Maniatis, and S.~Ratnasamy, ``Not-a-bot:
  Improving service availability in the face of {Botnet} attacks,'' in
  \emph{Proceedings of the 6th USENIX Symposium on Networked Systems Design and
  Implementation}, ser. NSDI'09.\hskip 1em plus 0.5em minus 0.4em\relax
  Berkeley, CA, USA: USENIX Association, 2009, pp. 307--320.

\bibitem{Fidelius}
S.~Eskandarian, J.~Cogan, S.~Birnbaum, P.~C.~W. Brandon, D.~Franke, F.~Fraser,
  G.~Garcia, E.~Gong, H.~T. Nguyen, T.~K. Sethi \emph{et~al.}, ``Fidelius:
  Protecting user secrets from compromised browsers,'' in \emph{2019 IEEE
  Symposium on Security and Privacy (SP)}.\hskip 1em plus 0.5em minus
  0.4em\relax IEEE, 2019, pp. 264--280.

\bibitem{context_hiding_attack}
Y.~Fratantonio, C.~Qian, S.~P. Chung, and W.~Lee, ``Cloak and dagger: from two
  permissions to complete control of the {UI} feedback loop,'' in \emph{2017
  IEEE Symposium on Security and Privacy (SP)}.\hskip 1em plus 0.5em minus
  0.4em\relax IEEE, 2017, pp. 1041--1057.

\bibitem{cleartype}
Microsoft, ``Microsoft cleartype overview,'' 2022,
  \url{https://docs.microsoft.com/en-us/typography/cleartype/} [Accessed Nov 26
  2022].

\bibitem{pixel_perfect}
K.~Mowery and H.~Shacham, ``Pixel perfect: Fingerprinting canvas in {HTML5},''
  in \emph{Proceedings of W2SP 2012}, M.~Fredrikson, Ed.\hskip 1em plus 0.5em
  minus 0.4em\relax IEEE Computer Society, May 2012.

\bibitem{image_hash_tampering}
S.~Roy and Q.~Sun, ``Robust hash for detecting and localizing image
  tampering,'' in \emph{2007 IEEE International Conference on Image
  Processing}, vol.~6.\hskip 1em plus 0.5em minus 0.4em\relax IEEE, 2007, pp.
  VI--117.

\bibitem{papernot2018cleverhans}
N.~Papernot, F.~Faghri, N.~Carlini, I.~Goodfellow, R.~Feinman, A.~Kurakin,
  C.~Xie, Y.~Sharma, T.~Brown, A.~Roy, A.~Matyasko, V.~Behzadan,
  K.~Hambardzumyan, Z.~Zhang, Y.-L. Juang, Z.~Li, R.~Sheatsley, A.~Garg,
  J.~Uesato, W.~Gierke, Y.~Dong, D.~Berthelot, P.~Hendricks, J.~Rauber, and
  R.~Long, ``Technical report on the {CleverHans} v2.1.0 adversarial examples
  library,'' \emph{arXiv preprint arXiv:1610.00768}, 2018.

\bibitem{croce2020reliable}
F.~Croce and M.~Hein, ``Reliable evaluation of adversarial robustness with an
  ensemble of diverse parameter-free attacks,'' in \emph{Proceedings of the
  37th International Conference on Machine Learning}, ser. ICML'20.\hskip 1em
  plus 0.5em minus 0.4em\relax JMLR.org, 2020.

\bibitem{clickshield}
A.~Possemato, A.~Lanzi, S.~P.~H. Chung, W.~Lee, and Y.~Fratantonio,
  ``Clickshield: Are you hiding something? towards eradicating {Clickjacking}
  on {Android},'' in \emph{Proceedings of the 2018 ACM SIGSAC Conference on
  Computer and Communications Security (CCS)}.\hskip 1em plus 0.5em minus
  0.4em\relax ACM, 2018, pp. 1120--1136.

\bibitem{ViperSof64:online}
J.~Rubin, ``Vipersoftx: Hiding in system logs and spreading venomsoftx - avast
  threat labs,''
  \url{https://decoded.avast.io/janrubin/vipersoftx-hiding-in-system-logs-and-spreading-venomsoftx/}
  [Accessed Feb 20 2023].

\bibitem{chen2008overshadow}
X.~Chen, T.~Garfinkel, E.~C. Lewis, P.~Subrahmanyam, C.~A. Waldspurger,
  D.~Boneh, J.~Dwoskin, and D.~R. Ports, ``Overshadow: a virtualization-based
  approach to retrofitting protection in commodity operating systems,''
  \emph{ACM SIGOPS Operating Systems Review}, vol.~42, no.~2, pp. 2--13, 2008.

\bibitem{kapravelos2014hulk}
A.~Kapravelos, C.~Grier, N.~Chachra, C.~Kruegel, G.~Vigna, and V.~Paxson,
  ``Hulk: Eliciting malicious behavior in browser extensions,'' in \emph{23rd
  {USENIX} Security Symposium ({USENIX} Security 14)}, 2014, pp. 641--654.

\bibitem{InfoSecH86:online}
``Infosec handlers diary blog - sans internet storm center,''
  \url{https://isc.sans.edu/diary/Clickjacking+attacks+on+Facebook%27s+Like+plugin/8893}
  [Accessed Feb 28 2023].

\bibitem{Explaini11:online}
``Explaining the {``Don't Click''} clickjacking {Tweetbomb},''
  \url{https://softwareas.com/explaining-the-dont-click-clickjacking-tweetbomb/}
  [Accessed Feb 28 2023].

\bibitem{WhydoIge64:online}
Zerodha, ``Why do {I} get an {SMS} and email from {NSE}, {BSE} and {MCX} when
  {I} trade?''
  \url{https://support.zerodha.com/category/trading-and-markets/trading-faqs/articles/why-do-i-get-an-sms-and-email-from-nse-bse-when-i-trade}
  [Accessed Nov 26 2022].

\bibitem{TradingA12:online}
I.~Brokers, ``Trading assistant,''
  \url{https://guides.interactivebrokers.com/iphone/monitor/trading-assistant.htm}
  [Accessed Nov 26 2022].

\bibitem{juba2015principled}
B.~Juba, C.~Musco, F.~Long, S.~Sidiroglou-Douskos, and M.~C. Rinard,
  ``Principled sampling for anomaly detection.'' in \emph{Proceedings of the
  2015 Network and Distributed System Security (NDSS) Symposium}, 2015.

\bibitem{xmhf}
A.~Vasudevan, S.~Chaki, L.~Jia, J.~McCune, J.~Newsome, and A.~Datta, ``Design,
  implementation and verification of an extensible and modular hypervisor
  framework,'' in \emph{Proceedings of the 2013 IEEE Symposium on Security and
  Privacy}, ser. SP '13.\hskip 1em plus 0.5em minus 0.4em\relax Washington, DC,
  USA: IEEE Computer Society, 2013, pp. 430--444.

\bibitem{tan2011tpm}
H.~Tan, W.~Hu, and S.~Jha, ``A {TPM}-enabled remote attestation protocol
  ({TRAP}) in wireless sensor networks,'' in \emph{Proceedings of the 6th ACM
  workshop on Performance monitoring and measurement of heterogeneous wireless
  and wired networks}, 2011, pp. 9--16.

\bibitem{nunes2019vrased}
I.~D.~O. Nunes, K.~Eldefrawy, N.~Rattanavipanon, M.~Steiner, and G.~Tsudik,
  ``Vrased: A verified hardware/software co-design for remote attestation.'' in
  \emph{USENIX Security Symposium}, 2019, pp. 1429--1446.

\bibitem{eckel2020secure}
M.~Eckel, A.~Fuchs, J.~Repp, and M.~Springer, ``Secure attestation of
  virtualized environments,'' in \emph{ICT Systems Security and Privacy
  Protection: 35th IFIP TC 11 International Conference, SEC 2020, Maribor,
  Slovenia, September 21--23, 2020, Proceedings 35}.\hskip 1em plus 0.5em minus
  0.4em\relax Springer, 2020, pp. 203--216.

\bibitem{screenshot_sidechannel}
E.~Fernandes, Q.~A. Chen, J.~Paupore, G.~Essl, J.~A. Halderman, Z.~M. Mao, and
  A.~Prakash, ``Android {UI} deception revisited: Attacks and defenses,'' in
  \emph{International Conference on Financial Cryptography and Data
  Security}.\hskip 1em plus 0.5em minus 0.4em\relax Springer, 2016, pp. 41--59.

\bibitem{exposure1}
W.~von Hippel and C.~Hawkins, ``Stimulus exposure time and perceptual memory,''
  \emph{Perception {\&} Psychophysics}, vol.~56, no.~5, pp. 525--535, Sep 1994.

\bibitem{rendering_variation1}
P.~Laperdrix, W.~Rudametkin, and B.~Baudry, ``Beauty and the beast: Diverting
  modern web browsers to build unique browser fingerprints,'' in \emph{2016
  IEEE Symposium on Security and Privacy (SP)}.\hskip 1em plus 0.5em minus
  0.4em\relax IEEE, 2016, pp. 878--894.

\bibitem{7163061}
F.~Yamaguchi, A.~Maier, H.~Gascon, and K.~Rieck, ``Automatic inference of
  search patterns for taint-style vulnerabilities,'' in \emph{2015 IEEE
  Symposium on Security and Privacy (SP)}, 2015, pp. 797--812.

\bibitem{gyrus}
Y.~Jang, S.~P. Chung, B.~D. Payne, and W.~Lee, ``Gyrus: A framework for
  user-intent monitoring of text-based networked applications.'' in
  \emph{Proceedings of the 2014 Network and Distributed System Security
  Symposium (NDSS)}, 2014.

\bibitem{ocr_ad_blocking}
F.~Tram\`{e}r, P.~Dupr\'{e}, G.~Rusak, G.~Pellegrino, and D.~Boneh,
  ``Adversarial: Perceptual ad blocking meets adversarial machine learning,''
  in \emph{Proceedings of the 2019 ACM SIGSAC Conference on Computer and
  Communications Security}, ser. CCS ’19.\hskip 1em plus 0.5em minus
  0.4em\relax New York, NY, USA: Association for Computing Machinery, 2019, p.
  2005–2021.

\bibitem{DBLP:journals/corr/HeZRS15}
K.~He, X.~Zhang, S.~Ren, and J.~Sun, ``Deep residual learning for image
  recognition,'' in \emph{Proceedings of the IEEE conference on computer vision
  and pattern recognition}, 2016, pp. 770--778.

\bibitem{emnist}
G.~Cohen, S.~Afshar, J.~Tapson, and A.~van Schaik, ``Emnist: Extending mnist to
  handwritten letters,'' in \emph{2017 International Joint Conference on Neural
  Networks (IJCNN)}, 2017, pp. 2921--2926.

\bibitem{cifar_10_tech_report}
A.~Krizhevsky and G.~Hinton, ``Learning multiple layers of features from tiny
  images,'' 2009,
  \url{https://www.cs.toronto.edu/~kriz/learning-features-2009-TR.pdf}
  [Accessed April 20 2023].

\bibitem{material_icon}
G.~Fonts, ``\BIBforeignlanguage{en}{Google {Fonts}},'' 2021,
  \url{https://fonts.google.com/} [Accessed April 20 2023].

\bibitem{goodfellow2014explaining}
I.~J. Goodfellow, J.~Shlens, and C.~Szegedy, ``Explaining and harnessing
  adversarial examples,'' \emph{arXiv preprint arXiv:1412.6572}, 2014.

\bibitem{adversarial_examples2}
N.~Papernot, P.~McDaniel, and I.~Goodfellow, ``Transferability in machine
  learning: from phenomena to black-box attacks using adversarial samples,''
  \emph{arXiv preprint arXiv:1605.07277}, 2016.

\bibitem{noauthor_github_nodate}
aymericdamien, ``\BIBforeignlanguage{en}{{GitHub} -
  aymericdamien/{TensorFlow}-{Examples}: {TensorFlow} {Tutorial} and {Examples}
  for {Beginners} (support {TF} v1 \& v2)},'' 2020,
  \url{https://github.com/aymericdamien/TensorFlow-Examples} [Accessed April 20
  2023].

\bibitem{noauthor_github_nodate-1}
arconsis, ``\BIBforeignlanguage{en}{{GitHub} -
  arconsis/cifar-10-with-tensorflow2},'' 2019,
  \url{https://github.com/arconsis/cifar-10-with-tensorflow2} [Accessed Oct 16
  2022].

\bibitem{szegedy2013intriguing}
C.~Szegedy, W.~Zaremba, I.~Sutskever, J.~Bruna, D.~Erhan, I.~Goodfellow, and
  R.~Fergus, ``Intriguing properties of neural networks,'' \emph{arXiv preprint
  arXiv:1312.6199}, 2013.

\bibitem{madry2017towards}
A.~Madry, A.~Makelov, L.~Schmidt, D.~Tsipras, and A.~Vladu, ``Towards deep
  learning models resistant to adversarial attacks,'' \emph{arXiv preprint
  arXiv:1706.06083}, 2017.

\bibitem{DBLP:journals/corr/abs-2103-01946}
S.-A. Rebuffi, S.~Gowal, D.~A. Calian, F.~Stimberg, O.~Wiles, and T.~Mann,
  ``Fixing data augmentation to improve adversarial robustness,'' \emph{arXiv
  preprint arXiv:2103.01946}, 2021.

\bibitem{gu2014towards}
S.~Gu and L.~Rigazio, ``Towards deep neural network architectures robust to
  adversarial examples,'' \emph{arXiv preprint arXiv:1412.5068}, 2014.

\bibitem{metzen2017detecting}
J.~H. Metzen, T.~Genewein, V.~Fischer, and B.~Bischoff, ``On detecting
  adversarial perturbations,'' \emph{arXiv preprint arXiv:1702.04267}, 2017.

\bibitem{xu2017feature}
W.~Xu, D.~Evans, and Y.~Qi, ``Feature squeezing: Detecting adversarial examples
  in deep neural networks,'' \emph{arXiv preprint arXiv:1704.01155}, 2017.

\bibitem{DBLP:journals/corr/KurakinGB16}
A.~Kurakin, I.~J. Goodfellow, and S.~Bengio, ``Adversarial examples in the
  physical world,'' in \emph{Artificial intelligence safety and
  security}.\hskip 1em plus 0.5em minus 0.4em\relax Chapman and Hall/CRC, 2018,
  pp. 99--112.

\bibitem{DBLP:journals/corr/abs-1710-06081}
Y.~Dong, F.~Liao, T.~Pang, X.~Hu, and J.~Zhu, ``Discovering adversarial
  examples with momentum,'' \emph{arXiv preprint arXiv:1710.06081}, vol.~5,
  2017.

\bibitem{cw_adv_exp}
N.~Carlini and D.~Wagner, ``Towards evaluating the robustness of neural
  networks,'' in \emph{2017 IEEE Symposium on Security and Privacy (SP)}.\hskip
  1em plus 0.5em minus 0.4em\relax Ieee, 2017, pp. 39--57.

\bibitem{DBLP:journals/corr/abs-1907-02044}
F.~Croce and M.~Hein, ``Minimally distorted adversarial examples with a fast
  adaptive boundary attack,'' in \emph{International Conference on Machine
  Learning}.\hskip 1em plus 0.5em minus 0.4em\relax PMLR, 2020, pp. 2196--2205.

\bibitem{papernot2016distillation}
N.~Papernot, P.~McDaniel, X.~Wu, S.~Jha, and A.~Swami, ``Distillation as a
  defense to adversarial perturbations against deep neural networks,'' in
  \emph{2016 IEEE Symposium on Security and Privacy (SP)}.\hskip 1em plus 0.5em
  minus 0.4em\relax IEEE, 2016, pp. 582--597.

\bibitem{sadko_guide_2021}
Y.~Sadko, ``\BIBforeignlanguage{en}{Guide to 10 font characteristics and their
  use in design},'' jan 2021,
  \url{https://eugenesadko.medium.com/guide-to-10-font-characteristics-and-their-use-in-design-b0a07cc66f7}
  [Accessed Nov 24 2022].

\bibitem{gelderman1999short}
M.~Gelderman, ``A short introduction to font characteristics,'' \emph{TUGboat},
  June 1999.

\bibitem{elsayed2018adversarial}
G.~F. Elsayed, S.~Shankar, B.~Cheung, N.~Papernot, A.~Kurakin, I.~Goodfellow,
  and J.~Sohl-Dickstein, ``Adversarial examples that fool both computer vision
  and time-limited humans,'' \emph{arXiv preprint arXiv:1802.08195}, 2018.

\bibitem{quach2018debloating}
A.~Quach, A.~Prakash, and L.~Yan, ``Debloating software through piece-wise
  compilation and loading,'' in \emph{27th USENIX Security Symposium (USENIX
  Security 18)}, 2018, pp. 869--886.

\bibitem{loading_speed}
H.~A. Org, ``Report:loading speed,'' Nov 2020,
  \url{https://httparchive.org/reports/loading-speed\#fcp} [Accessed April 20
  2023].

\bibitem{frame_bustering1}
G.~Rydstedt, E.~Bursztein, D.~Boneh, and C.~Jackson, ``Busting frame busting: a
  study of {Clickjacking} vulnerabilities at popular sites,'' \emph{IEEE
  Oakland Web}, vol.~2, no.~6, 2010.

\bibitem{zhao2009botgraph}
Y.~Zhao, Y.~Xie, F.~Yu, Q.~Ke, Y.~Yu, Y.~Chen, and E.~Gillum, ``Botgraph: large
  scale spamming botnet detection.'' in \emph{Proceedings of the 6th USENIX
  Symposium on Networked Systems Design and Implementation}, ser. NSDI'09,
  vol.~9, 2009, pp. 321--334.

\bibitem{von2008recaptcha}
L.~Von~Ahn, B.~Maurer, C.~McMillen, D.~Abraham, and M.~Blum, ``{reCAPTCHA}:
  Human-based character recognition via web security measures,''
  \emph{Science}, vol. 321, no. 5895, pp. 1465--1468, 2008.

\bibitem{zeller2008cross}
W.~Zeller and E.~W. Felten, ``Cross-site request forgeries: Exploitation and
  prevention,'' \emph{The New York Times}, pp. 1--13, 2008.

\bibitem{SameSite72:online}
MDN, ``Samesite cookies - http | mdn,'' May 2022,
  \url{https://developer.mozilla.org/en-US/docs/Web/HTTP/Headers/Set-Cookie/SameSite}
  [Accessed Nov 26 2022].

\bibitem{trusted_display}
M.~Yu, V.~D. Gligor, and Z.~Zhou, ``Trusted display on untrusted commodity
  platforms,'' in \emph{Proceedings of the 22Nd ACM SIGSAC Conference on
  Computer and Communications Security}, ser. CCS '15.\hskip 1em plus 0.5em
  minus 0.4em\relax New York, NY, USA: ACM, 2015, pp. 989--1003.

\bibitem{trusted_path}
Z.~Zhou, V.~D. Gligor, J.~Newsome, and J.~M. McCune, ``Building verifiable
  trusted path on commodity x86 computers,'' in \emph{Proceedings of the 2012
  IEEE Symposium on Security and Privacy}, ser. SP '12.\hskip 1em plus 0.5em
  minus 0.4em\relax USA: IEEE Computer Society, 2012, p. 616–630.

\bibitem{goltzsche2017trustjs}
D.~Goltzsche, C.~Wulf, D.~Muthukumaran, K.~Rieck, P.~Pietzuch, and R.~Kapitza,
  ``Trustjs: Trusted client-side execution of {JavaScript},'' in
  \emph{Proceedings of the 10th European Workshop on Systems Security}, 2017,
  pp. 1--6.

\bibitem{Shuang}
H.~Shuang, W.~Huang, P.~Bettadpur, L.~Zhao, I.~Pustogarov, and D.~Lie, ``Using
  inputs and context to verify user intentions in internet services,'' in
  \emph{Proceedings of the 10th ACM SIGOPS Asia-Pacific Workshop on Systems},
  ser. APSys '19.\hskip 1em plus 0.5em minus 0.4em\relax New York, NY, USA:
  ACM, 2019, pp. 76--83.

\bibitem{binder}
W.~Cui, R.~H. Katz, and W.-t. Tan, ``Binder: An extrusion-based break-in
  detector for personal computers,'' in \emph{Proceedings of the 2005 USENIX
  Annual Technical Conference}.\hskip 1em plus 0.5em minus 0.4em\relax
  Berkeley, CA, USA: USENIX Association, April 2005.

\bibitem{utp}
A.~Filyanov, J.~M. McCuney, A.-R. Sadeghiz, and M.~Winandy, ``Uni-directional
  trusted path: Transaction confirmation on just one device,'' in
  \emph{Proceedings of the 2011 IEEE/IFIP 41st International Conference on
  Dependable Systems and Networks}, ser. DSN '11.\hskip 1em plus 0.5em minus
  0.4em\relax Washington, DC, USA: IEEE Computer Society, 2011, pp. 1--12.

\bibitem{aware}
G.~Petracca, A.-A. Reineh, Y.~Sun, J.~Grossklags, and T.~Jaeger, ``Aware:
  Preventing abuse of privacy-sensitive sensors via operation bindings,'' in
  \emph{26th USENIX Security Symposium (USENIX Security 17)}, 2017, pp.
  379--396.

\end{thebibliography}

\end{document}